\documentclass[amssymb,aps,twocolumn,showpacs, superscriptaddress]{revtex4-1}
\usepackage[]{graphicx}
\usepackage[]{amsmath}
\usepackage{hyperref}
\usepackage{color}
\usepackage{xspace}
\usepackage{hyperref}

\def\ket#1{| #1 \rangle}
\def\bra#1{\langle #1 |}

\def\cO{\mathcal{O}}

\def\cU{\mathcal{U}}

\def\pr{{\rm Prob}}

\def\eq#1{Eq.~\eqref{eq:#1}}
\def\fig#1{Fig.~\ref{fig:#1}}
\def\sec#1{Sec.~\ref{sec:#1}}

\newcommand{\steps}{K}
\newcommand{\be}{\begin{equation}}
\newcommand{\ee}{\end{equation}}

\newcommand{\lqd}{LIQ$Ui|\rangle$\xspace}
\begin{document}

\title{The Trotter Step Size Required for Accurate Quantum Simulation of Quantum Chemistry}
\author{David Poulin}
\affiliation{D\'epartement de Physique, Universit\'e de Sherbrooke, Qu\'ebec, Canada}

\author{M. B. Hastings}
\affiliation{Station Q, Microsoft Research, Santa Barbara, CA 93106-6105, USA}
\affiliation{Quantum Architectures and Computation Group, Microsoft Research, Redmond, WA 98052, USA}

\author{Dave Wecker}
\affiliation{Quantum Architectures and Computation Group, Microsoft Research, Redmond, WA 98052, USA}

\author{Nathan Wiebe}
\affiliation{Quantum Architectures and Computation Group, Microsoft Research, Redmond, WA 98052, USA}

\author{Andrew C. Doherty}
\affiliation{Centre for Engineered Quantum Systems, School of Physics,\\
The University of Sydney, Sydney, NSW 2006, Australia}

\author{Matthias Troyer}
\affiliation{Theoretische Physik, ETH Zurich, 8093 Zurich, Switzerland}
\date{\today}

\begin{abstract}
The simulation of molecules is a widely anticipated application of quantum computers. However, recent studies \cite{WBCH13a,HWBT14a} have cast a shadow on this hope by revealing that the complexity in gate count of such simulations increases with the number of spin orbitals $N$ as $N^8$, which becomes prohibitive even for molecules of modest size $N\sim 100$. This study was partly based on a scaling analysis of the Trotter step required for an ensemble of random artificial molecules. Here, we revisit this analysis and find instead that the scaling is closer to $N^6$ in worst case for real model molecules we have studied, indicating  that the random ensemble fails to accurately capture the statistical properties of real-world molecules. Actual scaling may be significantly better than this due to averaging effects.  We then present an alternative simulation scheme and show that it can sometimes outperform existing schemes, but that this possibility depends crucially on the details of the simulated molecule. We obtain further improvements using a version of
the coalescing scheme of \cite{WBCH13a}; this scheme is based on using different Trotter steps for different terms.
The method we use to bound the complexity of simulating a given molecule is efficient, in contrast to the approach of \cite{WBCH13a,HWBT14a} which relied on exponentially costly classical exact simulation.
\end{abstract}


\maketitle

\section{Introduction}

It has been 30 years since Feynman suggested that a quantum information processor could in principle simulate the dynamics of quantum systems efficiently \cite{Fey82a}, and this idea has since been formalized and studied in great detail \cite{Llo96b,Zal98a,OGK+01a,AT03b,WBHS11a,BC12a,BCCK13a}. Based on this knowledge, it has been advocated that one of the first practical applications of quantum information processors will be the simulation of molecules \cite[\& references therein]{ADLH05a,WBA10a,KWPY11a}. This is motivated by the fact that state-of-the-art, high-precision numerical simulations are limited to molecules with at most  50-70 spin orbitals, where a spin orbital denotes a choice of both orbital and spin quantum numbers\cite{GH05a,NS11a,LSS11a,CDHL13a,KCY13a}. Thus, a quantum computer using as little as 100 logical qubits has enough storage capacity to efficiently perform a simulation which is otherwise intractable classically. 

However, closer scrutiny of the problem has recently revealed that, while the memory requirements are indeed relatively modest, the duration of such quantum simulations using the proposed techniques are far too demanding \cite{WBCH13a}. Significant improvements were obtained by optimizing the quantum simulation circuitry \cite{HWBT14a}, but the required time-resources remain prohibitive.  

To understand the origin of this problem, recall that the time-evolution operator associated to a Hamiltonian $H$ is $U_H(t) = e^{-iHt}$. For a Hamiltonian expressed as the sum of $m$ terms $H = \sum_{\alpha=1}^m H_\alpha$, we can use the Trotter-Suzuki (TS) decomposition to approximate the ``infinitesimal'' time-evolution operator $U_H(\Delta_t)$  by a product of $m$ infinitesimal time-evolution operators $U_\alpha(\Delta_t) = e^{-iH_\alpha\Delta_t}$, each generated by a single term $H_\alpha$ from the Hamiltonian. Repeating $1/\Delta_t$ times yields the time-evolution operator for a unit time. We can deduce two immediate consequences of this approach. On the one hand, the number of gates $N_g$ required to implement a single infinitesimal time step will scale at least proportionally to the number of terms $m$ in the Hamiltonian. On the other hand, the error in the TS approximation also increases as some power of $m$, forcing us to adopt a smaller time step $\Delta_t$, and hence a slower simulation \cite{WBHS11a,BCCK13a,WBCH13a}.

In the case of a molecule, the Coulomb force generates quartic terms $c_p^\dagger c^\dagger_q c_r c_s$ in fermion creation and annihilation operators. For small molecules, there can be as many as $\sim N^4$ such distinct terms, where $N$ is the number of relevant spin orbitals of the molecule.
The standard approach to the problem \cite{ADLH05a,WBA10a,KWPY11a,WBCH13a,HWBT14a} applies the TS decomposition directly to these $m=\cO(N^4)$ terms, each of which can be implemented using (at best) a constant number of gates\cite{HWBT14a} on average given a gate set containing one- and two-qubit Clifford operations as well as arbitrary single-qubit controlled rotations. Thus, this technique unavoidably entails at least a $\sim N^4$ cost per infinitesimal time-step.

Furthermore, to achieve a constant accuracy, the time evolution operator needs to be broken into a number of steps which increases as some power of $N$. In \cite{WBCH13a}, a rigorous upper bound on the TS error was derived which indicates that $1/\Delta_t = \cO(N^5)$ infinitesimal time-steps are sufficient to achieve a constant accuracy. This scaling was confronted with exact numerical simulations which revealed that $1/\Delta_t = \cO(N^{4-5})$ infinitesimal time-steps were indeed sufficient, resulting in a total complexity of $\cO(N^8)$ at best. This result is already exorbitant for a molecule with $N\sim100$ spin orbitals. 

These exact simulations were not carried on real-world molecules, but instead used artificial molecules drawn from a random ensemble meant to reproduce the statistical properties of real molecules. This is motivated by the fact that exact numerical simulations are restricted to small molecules $N\sim 24$ and the limited availability of interesting real-world molecules of small size. 

In this article, we assess the complexity of a quantum simulations without resorting to costly exact simulations, but instead directly and efficiently evaluate an upper bound derived in \cite{WBCH13a}. Because quantum simulations will be used precisely for those molecules that are too large to be amenable to classical simulations, this efficient and rigorous error assessment is also of independent interest. In this way, we are able to predict the complexity for the quantum simulations of real-world molecules of size up to $N\sim 100$, and find that $1/\Delta_t = \cO(N^{1.5-2.5})$ time-steps are sufficient to achieve a constant-accuracy simulation, and the true cost may be even lower. This result clearly indicates that the statistical properties of those molecules are not accurately reproduced by the random ensemble used in \cite{WBCH13a}, and that the complexity of simulating real-world molecules is substantially lower than anticipated.
Finally, we show that by using different TS steps for different terms, in a version of the coalescing approach of Ref.~\onlinecite{WBCH13a}, it is possible to obtain further improvements to the time complexity.

We also propose an alternative simulation scheme based on a decomposition of the Hamiltonian into a sum of $m=\cO(N^2)$ terms, each of which can be implemented with $\cO(N^2)$ gates. While this leads to an identical gate count $N_g \sim N^4$ per infinitesimal time step, reducing the number of terms in the TS decomposition can significantly reduce the resulting error, thus enabling a larger time step $\Delta_t$. We find that this alternative simulation scheme can sometimes outperform existing schemes, but that the performance of each scheme depends greatly on the details of the simulated molecule. Our efficient error assessment technique comes in handy at this point because it enables us to determine which simulation technique is best suited for a given molecule. This illustrates that other decompositions of the Hamiltonians could lead to substantial gains.

In general, in this paper when estimating work we will count the number of {\it gates} required.  The nesting scheme of Ref.~\onlinecite{HWBT14a} means that in many cases we can parallelize such that the {\it depth} of the circuit will be proportional to the number of gates divided by $N$.  In a few places we comment on this more explicitly.

\section{Background}
\label{sec:background}

In this section, we present the problem more formally, and review the standard simulation approach \cite{ADLH05a,WBA10a,KWPY11a,WBCH13a,HWBT14a}.

\subsection{Hamiltonian}
Our starting point is a Hamiltonian of the form
\begin{equation}
H = \sum_{pq} h_{pq} c_p^\dagger c_q + \sum_{pqrs} h_{pqrs} c^\dagger_pc_q^\dagger c_r c_s,
\label{eq:H}
\end{equation}
where $c^\dagger_p$ and $c_p$ are fermion creation and annihilation operators for the spin orbital $p$. There are $N$ spin orbitals which have been chosen using, e.g. Hartree-Fock calculations. To get a constant-accuracy estimate of the ground-state energy of the corresponding molecule, we need to simulate the time-evolution operator $U_H(t)$ for some constant time $t$, which we will set to unity in what follows and drop the explicit $t$ variable when unnecessary. 

The first term of \eq{H} describes free fermions. We will use the shorthand notation $H_{pq} = h_{pq} c^\dagger_pc_q$ for these terms and note that $H_{pq}^\dagger = H_{qp} \Leftrightarrow h_{pq}^* = h_{qp}$ is required for $H$ to be Hermitian. A nice property of free-fermion operators is that they form a closed Lie algebra, i.e., for $H = \sum_{pq} h_{pq} c^\dagger_p c_q$ and $H' = \sum_{pq} h'_{pq} c^\dagger_p c_q$, we have 
\begin{equation}
[H,H'] = \sum_{pq} [h,h']_{pq} c^\dagger_p c_q,
\label{eq:comm}
\end{equation} 
where $[h,h']_{pq}$ simply refers to the $(p,q)$ matrix element of $[h,h']$.  Throughout, we will use upper-case letters $H,U,\ldots$ to denote operators on the $2^N$-dimensional Hilbert space, and lower-case letters $h,u,\ldots$ for matrices on the $N$-dimensional orbital space. The group $\cU(N)$ acts on the orbital space, and \eq{comm} simply shows that free fermion Hamiltonians form a (reducible) representation of $\cU(N)$. In other words, $e^{-iHt}c_pe^{iHt} = \sum_q u_{pq}c_q$ where $u = e^{-iht}\in \cU(N)$. 

The second term of \eq{H} represents interactions. We will use the shorthand notation $H_{pqrs} = h_{pqrs}c^\dagger_pc_q^\dagger c_rc_s$. We note that the substitution $h_{pqrs} \leftarrow \frac{h_{pqrs}+h_{qpsr}}2$ leaves the Hamiltonian invariant, so we will henceforth assume that $h_{pqrs} = h_{qpsr}$. 

With the exception of section \ref{sec:simul2}, where all four fermion terms are considered on an equal footing, we will reserve $H_{pqrs}$ to refer to terms where $p,q,r,s$ are all distinct and otherwise refer to $H_{prrq}$ terms and $H_{pqqp}$ terms to refer to the case that only $3$ or $2$ of the indices are distinct.  Note that the terms $H_{pqqp}$ are diagonal in an occupation number basis.  Similarly, we use $H_{pp}$ to refer to terms proportional to $c^\dagger_p c_p$ and $H_{pq}$ to refer to terms proportional to $c^\dagger_p c_q$ for $p\neq q$.

\subsection{Simulating time evolution}
\label{sec:evolution1}
The general strategy to simulate the time-evolution generated by a Hamiltonian which is the sum of $m$ simple terms $H = \sum_{\alpha=1}^m H_\alpha$ proceeds in two phases. First, we decompose the total time evolution into a sequence of infinitesimal steps $U_H(1) = [U_H(\Delta_t)]^{1/\Delta_t}$. Second, we use the second-order (or higher) TS decomposition to approximate each infinitesimal steps
\begin{align}
&U_H(\Delta_t) \approx U_H^{\rm TS}(\Delta_t)\\
&:= U_m(\tfrac{\Delta_t}2)\ldots U_2(\tfrac{\Delta_t}2)U_1({\Delta_t})U_2(\tfrac{\Delta_t}2)\ldots U_m(\tfrac{\Delta_t}2),
\label{eq:TS}
\end{align}
where  $U_\alpha(t) = e^{-iH_\alpha t}$. For the Hamiltonian of \eq{H}, the index $\alpha$ would range over all the pairs $\alpha = (p,q)$ and quartets $\alpha = (p,q,r,s)$ of spin orbitals, so $m\sim N^4$. 

We claim that the evolution generated by any free Hamiltonian  can be implemented exactly with $\cO(N^2)$ gates. This follows from the Householder transformation \cite{HJ85a} which shows that we can decompose $u = e^{-ih} \in \cU(N)$ into a sequence of $N^2$ unitary matrices $u = v_1v_2\ldots v_{N^2}$, where each $v_\alpha$ is trivial everywhere except on a $2\times 2$ block $(p_k,q_k)$. Writing $v_k = e^{-ig^k}$, we can express  $U_H = V_1V_2\ldots V_{N^2}$ where $V_k = \exp(-ig^k_{p_k q_k} c^\dagger_{p_k}c_{q_k} + h.c.)$ is a time-evolution operator generated by a free-fermion operator acting only on 2 modes.  These operators $V_k$ can be implemented with a constant number of gates on average using the technique of Ref.~\cite{HWBT14a} to cancel Jordan-Wigner strings\footnote{This constant average cost can only be achieved when an appropriate sequence of free evolution operators are executed, which will always be the case here. Otherwise the cost is $\sim N$ due to the gates that are required to implement the Jordan-Wigner mapping of fermion orbitals into qubits.}. We note that this simulation of free Hamiltonians is exact, in contrast to previous approaches \cite{ADLH05a,WBA10a,KWPY11a,WBCH13a,HWBT14a} that rely on TS approximations. This is crucial for the alternative simulation scheme we will present in \sec{simul2} because it makes frequent uses of such free evolution operators to implement spin orbital basis changes.  
Further, it is possible to choose the ordering of the free-fermion evolution operators such that nesting as in Ref.~\onlinecite{HWBT14a} can be used to reduce the depth to $\cO(N)$.

Simulating the interaction terms is more demanding. In Ref.~\onlinecite{HWBT14a}, it was shown how the time evolution generated by each term $H_{pqrs}$ can be implemented using a constant number of gates on average. Since there are far more interaction terms than free terms, the overall circuit complexity is set by them, and the number of gates $N_g$ required to implement a single infinitesimal time-evolution operator \eq{TS} is therefore $\sim N^4$.

In practice, although it is possible to simulate all the free fermion terms in Eq.~(\ref{eq:H}) exactly without any TS error as described above, it would likely be preferred to use the scheme of Ref.~\onlinecite{HWBT14a} in which these terms are interleaved with terms $H_{prrq}$ so that after a term $c^\dagger_p c_q$ is executed, it is followed by terms $c^\dagger_p c^\dagger_r c_r c_q$.  In this way, in a Hartree-Fock basis these terms tend to cancel each other, reducing the TS error.

\subsection{Error per infinitesimal time-step}

Following the analysis of \cite[appendix B]{WBCH13a}, the approximation in \eq{TS} results in an error bounded by 
\begin{align}
&\delta^{\rm TS} := \|U_H(\Delta_t)-U_H^{\rm TS}(\Delta_t)\| \\
&\leq \sum_{\alpha=1}^m \Bigg\| [[H_\alpha,H_{>\alpha}],H_\alpha]] + [[H_{>\alpha},H_\alpha],H_{>\alpha}]]  \Bigg\| \Delta_t^3\label{eq:bound}
\end{align}
where $H_{>\alpha} = \sum_{\beta > \alpha}H_\beta$. Note that $[H_{pqrs},H_{p'q'r's'}] = 0$ unless one of the indices is repeated. Thus, of all the $m^3 = \cO(N^{12})$ terms $[[H_\alpha,H_\beta],H_\gamma]$ appearing in \eq{bound}, only $mK^2$ will be on-zero, where $K = \cO(N^3)$ is the maximum number of terms $H_\beta$ with which a given $H_\alpha$ does not commute. Defining $\Lambda = \max_\alpha \|H_\alpha\|$, which is a constant in the present case, this leads to the upper bound
\begin{equation}
\delta^{\rm TS}
\leq mK^2\Lambda^3 \Delta_t^3 = \cO(N^{10})\Delta_t^3.
\label{eq:TSerror1}
\end{equation}

The error in TS evolution gives an upper bound to the error in the eigenvalues of the unitary operator to evolve for a time step $\Delta_t$.
This translates to an error in the ground state energy
\begin{eqnarray}
\Delta E^{\rm TS}
&\leq &\sum_{\alpha=1}^m \Bigg\| [[H_\alpha,H_{>\alpha}],H_\alpha]] + [[H_{>\alpha},H_\alpha],H_{>\alpha}]]  \Bigg\| \Delta_t^2\nonumber \\
& =& \cO(N^{10})\Delta_t^2.
\label{eq:GSerror1}
\end{eqnarray}
This implies that the time step $\Delta_t$ needs to be as little as $\Delta_t = \cO(N^{-5})$ to achieve a constant precision, which is the rigorous upper bound derived in \cite{WBCH13a}.

An alternate route to undertanding error is to use the Baker-Campbell-Hausdorff (BCH) formula to compute the error in the TS approximation as a power series in $\Delta_t$.
Remarkably, the lowest order term in the power series gives an error that is within a constant factor of that resulting from the bound above, implying that the bound is close to optimum.
The advantage of the BCH formula is that it gives a tighter estimate of error for small $\Delta_t$.  The advantage of the bound above is that it works for all $\Delta_t$ while using the BCH formula it would be necessary also to consider higher-order terms in the power series.

The accuracy of the upper bounds in~\eqref{eq:bound} can be assessed by comparing the values it predicts to the asymptotic formula for the error.  The BCH formula can provide this by giving the leading order behavior of the \emph{effective Hamiltonian} that the TS simulation evolves under.  Applying the formula iteratively to~\eqref{eq:TS} to order $\Delta t^3$ yields
\begin{align}
H_{\rm eff}\! =\!H\!-\! \frac{1}{12}\! \sum_{\alpha\le \beta}\!\!\sum_{\beta}\!\!\sum_{\alpha'< \beta}\!\! \big[\!H_\alpha(1\!-\!\frac{\delta_{\alpha,\beta}}{2}),\!\big[\!H_\beta,\!H_{\alpha'}\big]\!\big]\!\Delta_t^2\label{eq:Heff}.
\end{align}
The error in the TS expansion for a single time step is therefore \begin{equation}
\|e^{-i H \Delta_t} - e^{-i H_{\rm eff} \Delta_t}\|\le \|H-H_{\rm eff}\|\Delta_t.\label{eq:standarderrbd}
\end{equation} Similarly, perturbation theory gives that the error in the ground state energy invoked by using the TS formula is
\begin{equation}
\! \sum_{\alpha\le \beta}\!\!\sum_{\beta}\!\!\sum_{\alpha'< \beta}\!\!
 \frac{1}{12}\bra{\Psi_0} \big[H_\alpha(1\!-\!\frac{\delta_{\alpha,\beta}}{2}),\!\big[H_\beta,H_{\alpha'}\big]\!\big]\!\Delta_t^2\ket{\Psi_0},\label{eq:HeffErr}
\end{equation}
where $\ket{\Psi_0}$ is the ground state of the true Hamiltonian $H$ and terms of order $\cO(\Delta_t^3)$ have been neglected. These formulas are valuable because they exactly predict the errors in the simulation as $\Delta_t$ approaches zero.

 If we apply the triangle inequality to~\eqref{eq:standarderrbd} then we obtain a comparable result to~\eqref{eq:bound} to within roughly a factor of $12$ if we neglect the $\cO(\Delta t^3)$ terms.  This means that~\eqref{eq:bound} can be expected to be reasonably tight in the regime where the error scales proportional to $\Delta_t^2$, up to errors incurred by using the triangle inequality.

\subsection{Random ensemble}
While the bound derived in the previous section is rigorous, it is possible that the  actual accuracy achieved in a quantum simulation is much better.  In Ref.~\onlinecite{WBCH13a}, this question was addressed using full classical simulations of the quantum simulation algorithm itself. While such a numerical simulation are certainly not tractable (this is the entire point of using a quantum information processor), they can be realized for molecules of modest sizes, and this can provide an idea of the general scaling. Due to the limited availability of interesting real-world molecules of small size, the simulations of \cite{WBCH13a} were carried on artificial molecules drawn from a random ensemble. 

Specifically, for Hamiltonians of this ensemble, only a fraction $F \approx 0.8\%$ of the entries $h_{pqrs}$ of the Hamiltonian are assigned non-zero values, those non-zero values have random signs and magnitudes following the distribution
\begin{align}
\pr(|h_{pqqp}|) &= {\rm Uniform}(0,0.5) \label{eq:E1}\\
\pr(|h_{pqqr}|) &= {\rm Exponential}(0.2) \label{eq:E2} \\
\pr(|h_{pqrs}|) &= {\rm Exponential}(0.1) .\label{eq:E3}
\end{align}
These simulations revealed that $\Delta_t \sim N^{-4.32}$-$N^{-5.08}$ depending on the electronic filling factor. In our numerical studies, we have also considered the ensemble obtained with $F=1$.

\section{Efficiently computable error bound}
\label{sec:BetterErr}

There is an obvious way of improving the upper-bound derived above using efficient numerical calculations. First, notice that the first term $[[H_\alpha,H_{>\alpha}],H_\alpha]$ of \eq{bound} contains only $K$ terms in contrast to the second term $[[H_\alpha,H_{>\alpha}],H_\alpha]$ which contains $K^2$. Thus, we will henceforth drop the first term for simplicity. Then, using the Jacobi identity $[[A,B],C] = [A,[B,C]] - [B,[A,C]]$, we see that $[[H_\alpha,H_\beta],H_{\beta'}]$ is zero unless two of the three pairs of terms from $H_\alpha$, $H_\beta$ and $H_{\beta'}$ do not commute.  Combined with the triangle inequality, we obtain the following upper bound to the second term of \eq{bound}
\begin{equation}
\delta^{\rm TS} \leq 4 \sum_\alpha \|H_\alpha\| \Bigg( \sum_{\beta}^{}\!{}' \|H_\beta\|\Bigg)^2\Delta_t^3,
\label{eq:TSerror2}
\end{equation}
where $\Sigma'$ is the sum restricted to the terms $\beta$ for which $[H_\alpha,H_\beta]\neq 0$. 
This gives an error in ground state energy 
\begin{equation}
\Delta E^{\rm TS} \leq 4 \sum_\alpha \|H_\alpha\| \Bigg( \sum_{\beta}^{}\!{}' \|H_\beta\|\Bigg)^2\Delta_t^2,
\end{equation}
In \eq{TSerror2}, we made double use of the inequality 
\begin{equation}
\|[H_\alpha,H_\beta]\|\leq 2\|H_\alpha\|\cdot\|H_\beta\|.
\label{eq:BoundComm}
\end{equation}
We note however that in the case where each term $H_\alpha$ represents a $H_{pqrs}$ term, this inequality is tight up to a factor of 2, provided that the terms do not commute. For instance, $\|[H_{pqrs},H_{p'q'p s'}]\| = |h_{pqrs}|\cdot |h_{p'q'ps'}|\cdot \|c_{p'}^\dagger c_{q'}^\dagger c_{s'}c_q^\dagger c_rc_s\| = |h_{pqrs}|\cdot |h_{p'pr's'}|$. 
 Thus, given a description of the molecule in terms of the $m$ coefficients $h_{pqrs}$, the bound \eq{TSerror2} can be evaluated with a complexity linear in $m$.
Since $m \sim N^4$, the evaluation of this bound could become numerically demanding for large molecules $N\gg 100$. Moreover, the evaluation of a similar bound for the alternative simulation approach of \sec{simul2} scales like $N^{9}$, so it becomes necessary to develop more efficiently ways of evaluating \eq{TSerror2}.  This can be achieved by Monte Carlo sampling from the sum rather than evaluating every terms. More precisely, we can generate $M$ triples of indices $\alpha_k$, $\beta_k$ and $\beta_k'$ such that both $[H_{\alpha_k},H_{\beta_k}]$ and $[H_{\alpha_k},H_{\beta_k'}]$ are non-zero, and estimate the bound in \eq{TSerror2} by
\begin{equation}
\label{triplets}
\delta^{\rm MC} =  \frac LM \sum_{k=1}^M |h_{\alpha_k}|\cdot |h_{\beta_k}|\cdot |h_{\beta'_k}|
\end{equation}
where $L$ is the total number of triplets $\alpha_k$, $\beta_k$ and $\beta_k'$ that obey the above conditions. The relative error on this estimate is $\sigma/\delta^{\rm MC}\sqrt M$, where $\sigma$ is the variance of $\delta^{\rm MC}$ and can also be estimated by sampling. In all the cases in which we were forced to use Monte Carlo sampling to estimate error upper bounds, we have used $M=10^5$ samples and observed that the relative error on our estimate of $\delta^{\rm MD}$ was about 1\% or less.  Moreover, Monte Carlo sampling is used only for the alternative simulation approach described in Sec.~\ref{sec:simul2}.

\section{Alternative decomposition}
\label{sec:simul2}

In this section, we present an alternative way of simulating molecules described by Hamiltonians of the form \eq{H} and show how to efficiently evaluate its accuracy.

\subsection{Completing the square} 

The scheme we propose is based on the idea of expressing the interacting Hamiltonian as the sum of squares of free terms plus additional free terms, i.e., completing the squares. We begin by demonstrating how this is realized. By joining indices $(p,r) = \alpha$ and $(q,s) = \beta$, we can view the tensor $h_{pqrs}$ as a matrix $h_{\alpha\beta}$. This matrix is symmetric given our convention $h_{pqrs} = h_{qpsr}$, so it can be diagonalized into $h_{\alpha\beta} = \sum_\gamma u_{\alpha\gamma} d_\gamma u^*_{\gamma\beta}$ with some unitary matrix $u$ and real vector $d$. 

Consider the operator $K_\gamma = \sum_{rs}u^*_{\gamma qs} c^\dagger_qc_s$, where we have partly converted back our notation $(q,s) = \beta$. The interaction Hamiltonian can now be expressed as $-\sum_{\gamma=1}^{N^2} d_\gamma K^\dagger_\gamma K_\gamma + H_1$, where $H_1 = \sum_{pqs} h_{pqqs}c^\dagger_p c_s$ is a quadratic term. While $K_\gamma$ are not Hermitian operators, they can be written as the sum of Hermitian and skew-Hermitian operators $K_\gamma = (\hat K_\gamma + i\tilde K_\gamma)$,  leading to $K^\dagger_\gamma K_\gamma = \hat K_\gamma^2 + \tilde K_\gamma^2 +i [\hat K_\gamma,\tilde K_\gamma]$. The commutator results in a free Hamiltonian. Defining $G_\gamma = \sqrt{|d_\gamma|}\hat K_\gamma$ and $G_{\gamma+N^2} = \sqrt{|d_\gamma|}\tilde K_\gamma$, we have expressed the Hamiltonian \eq{H} as a sum of squares of free Hamiltonians
\begin{equation}
H = H_0 - \sum_{\gamma =1}^{2N^2} \eta_\gamma G_\gamma^2
\label{eq:H2}
\end{equation} 
where $\eta_\gamma = \eta_{\gamma+N^2} = {\rm sign}(d_\gamma)$ and $H_0$ is a free Hamiltonians containing the initial free term of \eq{H}, the $H_1$ component above, plus the various commutators $i [\hat K_\gamma,\tilde K_\gamma]$. 

\subsection{Simulation}

Now that we have expressed the Hamiltonian in the form \eq{H2}, the simulation proceeds by using a second-order TS decomposition \eq{TS} as above, but this time using only $m=2N^2+1$ terms; $H_0$ and the $G_\gamma^2$. To implement an infinitesimal time evolution $U_\gamma=\exp(-i\eta_\gamma G_\gamma^2\Delta_t)$ generated by a $G_\gamma^2$ term, we first change the basis of orbitals so as to diagonalize this term. Given $G_\gamma = \sum_{pq} g_{pq}^\gamma c_p^\dagger c_q$, we can diagonalize the matrix $g^\gamma$ into $w^{\gamma}g^\gamma w^{\gamma\dagger} = \epsilon^\gamma $ where $\epsilon^\gamma$ is a diagonal matrix. Thus, written in the orbital basis $f_p^\gamma = \sum_q w^\gamma_{pq} c_q$, the term $G_\gamma = \sum _p \epsilon^\gamma_p f^{\gamma\dagger}_pf_p^\gamma$ is diagonal.  Using techniques of Ref.~\cite{HWBT14a}, it follows that in this orbital basis, the infinitesimal time evolution $U_\gamma$ can be realized using $\cO(N)$ gates.

The complexity of each $U_\gamma$ therefore stems from the $\cU(N)$ orbital basis change $w^{\gamma}$. But such a basis change is equivalent to time-evolution under a free Hamiltonian, so its complexity is $\cO(N^2)$ as explained in \sec{evolution1}. Thus, the overall complexity of simulating a single infinitesimal time evolution is the number of terms in the TS decomposition $\cO(N^2)$, times the complexity of implementing a single term $\cO(N^2)$,  resulting in the same scaling $\cO(N^4)$ as the method \cite{HWBT14a} outlined in \sec{evolution1}.

\subsection{Error per infinitesimal time-step}

We now evaluate the general error bound \eq{bound} in the special case where each term $H_\alpha = \pm G_\alpha^2$ is the square of a free fermion Hamiltonian. While we could proceed the same way as what led to \eq{TSerror2}, we note that the bound \eq{BoundComm} is not tight except in the special case explained in \sec{BetterErr}. So instead, we make use of the bound
\begin{equation}
[G_\alpha^2,G_\beta^2] \leq 4\|G_\alpha\|\cdot\|G_\beta\|\cdot \|[G_\alpha,G_\beta]\|,
\label{eq:comm_bound}
\end{equation}
which, inserted into in \eq{TSerror1} and combined with the triangle inequality, yields
\begin{equation}
\delta^{\rm TS} \leq 8  \sum_{\beta,\beta'>\alpha} \|G_\alpha\|\cdot \|G_\beta\|\cdot \|G_{\beta'}\|\cdot \|[[G_\alpha,G_\beta]G_{\beta'}] \| \Delta_t^3
\label{eq:TSerror3}
\end{equation}
Note that each of the terms $G_\alpha$, $G_\beta$, $G_{\beta'}$ and $[[G_\alpha,G_\beta],G_{\beta'}]$ are free fermion operators, so their norm can be computed efficiently numerically. Indeed, the operator norm  of a free Hamiltonian $H = \sum_{pq} h_{pq}c^\dagger_pc_q$ can be computed from the spectrum $\epsilon_p$ of the corresponding $h$ as $\|H\| = \max\{E_+,-E_-\}$, where $E_+$ ($E_-$) is the sum of the positive (negative) eigenvalues $\epsilon_p$. It follows that computing the norm of such an operator has complexity $\cO(N^3)$, and therefore evaluating the bound \eq{TSerror3} has overall complexity $\cO(N^9)$. For this reason, we have resorted to Monte Carlo sampling as explained in \sec{BetterErr} to evaluate \eq{TSerror3}.

 \section{Numerical results}
 \begin{figure}[t]
\includegraphics[width=8cm]{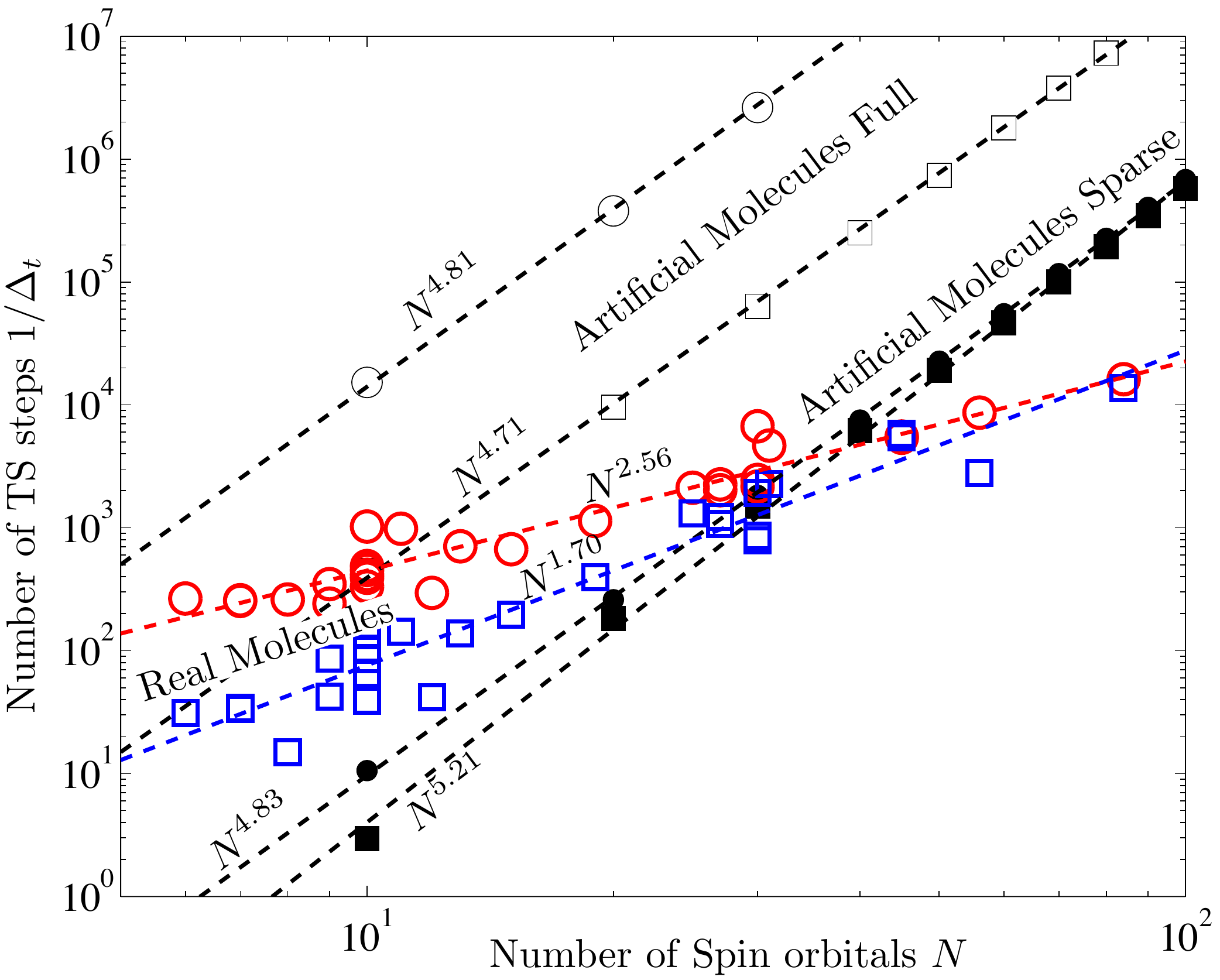}
\caption{Number of TS steps required to achieve constant accuracy on the energy measurement. Circles correspond to the standard approach described in \sec{background} and is derived from \eq{TSerror2}. Squares correspond to the  scheme described in \sec{simul2} and is derived from \eq{TSerror3}. Coloured marks are for a suite of real-world molecules. Filled black marks are for artificial molecules drawn from the sparse Hamiltonian ensemble $F=0.008$ and hollow marks are for artificial molecules drawn from the full Hamiltonian ensemble $F=1$.}
\label{fig:results}
\end{figure}

We have evaluated the bounds \eq{TSerror2} and \eq{TSerror3} associated to the two simulation schemes. Since $\delta^{\rm TS}$ is the error of a single infinitesimal TS time-step and that there are in total $1/\Delta_t$ such time-steps, a bound of the form $\delta^{\rm TS} \leq \Gamma \Delta_t^3$ implies that $\sqrt \Gamma$ time-steps are requires to achieve a constant accuracy. \fig{results} therefore presents the numerical values obtained for \eq{TSerror2} and \eq{TSerror3} in terms of the total number of TS steps $1/\Delta_t$. The bounds were evaluated for real-world molecules, for artificial molecules chosen from the ensemble described at Eqs.~(\ref{eq:E1}-\ref{eq:E3}), and for artificial molecules chosen from a different random ensemble where non-zero values were assigned to {\em all} $h_{pqrs}$ coefficients following the distribution Eqs.~(\ref{eq:E1}-\ref{eq:E3}). We refer to these two random ensembles as sparse and full respectively.

\subsection{Artificial molecules}

For the artificial molecules, we find that the fraction $F$ of non-zero coefficients $H_{pqrs}$ has little impact on the scaling of the complexity with $N$, and both simulation schemes display a complexity near $N^{4.5}$-$N^5$, in good agreement with the findings of \cite{WBCH13a} obtained from full numerical simulations. However, as we discuss below, this is likely an artifact of small sizes and the true scaling even for the artifical molecules is likely much better.

Although the scaling with $N$ is largely insensitive to the chosen random ensemble, we find that this choice greatly affects the constant pre factor. This is anticipated from the fact that a denser Hamiltonian will have a correspondingly higher norm, which will directly translate into a higher TS error bound. However, we observe that the two simulation schemes are not affected equally by this Hamiltonian density: while both schemes are have nearly identical complexity on the ensemble with $F=0.008$, the estimates for the scheme of \sec{simul2} are $100$ times better on the ensemble with $F=1$. 

\begin{figure}[t]
\includegraphics[width=8cm]{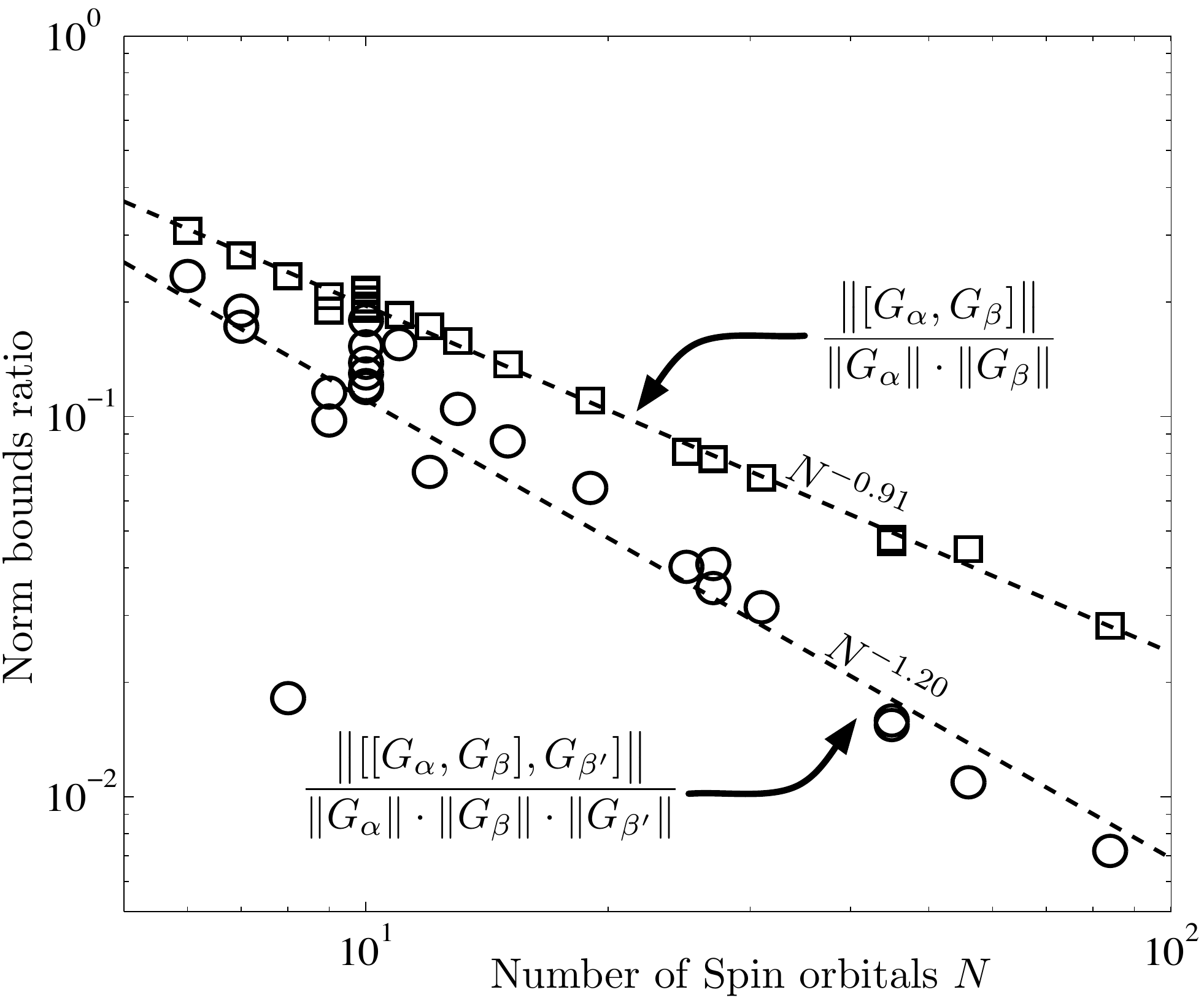}
\caption{Loss suffered from using bounds in \eq{BoundComm} instead of \eq{comm_bound} as a function of the number of spin orbitals.}
\label{fig:comm_bound}
\end{figure}

\subsection{Real molecules}

Despite the rather dispersed data, there appears to be a clear discrepancy with the results obtained from real and artificial molecules. This strongly suggests that the scaling with real molecules is much more favorable than the one anticipated from simulations of artificial molecules \cite{WBCH13a}, with a scaling in the range $N^{1.5}$-$N^{2.5}$ instead of $N^4$-$N^5$. The standard simulation scheme appears to offer a better scaling than the scheme of \sec{simul2}, but the data is too scattered to draw any firm conclusion. A case-by-case approach seems the most appropriate at this stage.

\subsection{Analysis}
\label{analysis}
The artificial molecule ensemble with $F=1$ illustrate that, by choosing to decompose the Hamiltonian \eq{H} into a sum of fewer but more complex terms \eq{H2}, we can obtained significant improvements of the TS error in our quantum simulation algorithm. It is unclear at this stage how much of this gain is real, and how much is coming from a tighter upper bound on the error. On the one hand, all these bounds make use of the triangle inequality to bound the sum of $M$ terms as follows
\begin{equation}
\Big\|\sum_{i=1}^MX_i\Big\| \leq \sum_{i=1}^M \|X_i\| \leq M \max_i\|X_i\|.
\end{equation}
This bound may not be tight, since we might expect the true norm  to scale  like $\sqrt M$ instead of $M$ due to some averaging effect.  However, surprisingly the agreement between the scaling of the bound agrees with the one found in \cite{WBCH13a} using exact simulations for the ensemble of
artificial molecules.  This initially
suggests that not much is lost in triangle inequalities.  We now analyze this in more detail using analytic estimates and additional simulations using
\lqd, a quantum simulator developed at Microsoft Research\cite{liquidref}, and argue that at the sizes of molecules amenable to simulation (including those in \cite{WBCH13a}) the terms arising from the commutator of distinct $H_{pqrs}$ terms are not yet important, but will become important at larger sizes. However, we further argue that the correct error anlysis will scale only as $\sqrt M$ due to average so that the true error estimate is significantly better than predicted by the triangle inequalities.

For this subsection, we primarily focus on the error estimates for the standard decomposition.
For the alternative decomposition of section \ref{sec:simul2}, another source of improvements comes from our ability to efficiently evaluate the operator norm of free fermion operators. For general operators $G_\alpha$, $G_\beta$, the norm of their commutator is bounded by \eq{BoundComm}. 
However, when $G_\alpha$, $G_\beta$ are free fermion operators, we can directly evaluate their commutator and its norm. This idea is used to derive \eq{TSerror3}, which should be much tighter than the corresponding naive upper bound, and could also partly explain the observed gain. \fig{comm_bound} illustrates the average advantage of evaluating the norm of the commutator instead of using a naive bound \eq{BoundComm} for the real molecules studied in \fig{results}. 

Consider the commutator  $[[H_{>\alpha},H_\alpha],H_{>\alpha}]]$ in Eq.~(\ref{eq:bound}).  Above, we used a triangle inequality to upper bound this
commutator by summing norms of double commutators $\Vert [[H_\alpha,H_\beta],H_\beta] \Vert$.  There are at most $\cO(N^{10})$ non-vanishing commutators, so that this estimate is at most $\cO(N^{10})$.  However, we also have available the bound 
that
$\Vert [[H_{>\alpha},H_\alpha],H_{>\alpha}]] \Vert \leq 4 \Vert H_{\alpha} \Vert \cdot \Vert H_{>\alpha} \Vert^2$.
For the artificial molecule ensembles with $h_{pqrs}$ assigned random signs, note that $H_{>\alpha}$ is a sum of $\cO(N^4)$ terms with uncorrelated random signs and magnitude of order unity.  Hence, we expect that it will have $\Vert H_{>\alpha} \Vert \leq \cO(N^2)$.
If this estimate holds, we have $\sum_{\alpha} \Vert [[H_{>\alpha},H_\alpha],H_{>\alpha}]] \Vert \leq \cO(N^8)$ which already improves on the $\cO(N^{10})$ estimate.  This estimate of $\cO(N^8)$ completely ignores any considerations of which terms in $H_{>\alpha}$ commute with $H_{\alpha}$ and hence is still likely to be an overestimate; perhaps improved estimates could be obtained based on applying the trace method directly to the double commutator 
$[[H_{>\alpha},H_\alpha],H_{>\alpha}]]$ as this double commutator is a sum of $\cO(N^6)$ terms with random signs but with some correlation between the signs.
However, the estimate $\cO(N^8)$ already hints that the upper bound from the triangle inequality is not tight.  In this subsection, we further explore the possibility that averaging improves these estimates (the estimate for $\Vert H_{\alpha} \Vert$ is an example of a kind of averaging as the norm of a sum of terms may be much less than the sum of the norms).  In an appendix, we briefly discuss the extent to which we can show the estimate 
$\Vert H_{>\alpha} \Vert \leq \cO(N^2)$.

In computing the error in numerical simulation, in all cases we chose the time step $\Delta_t$ sufficiently small to enter the regime that
error scaled proportional to $\Delta_t^2$.
The first piece of numerical evidence that the bounds are not yet relevant at the available $N$ is that an attempt to correlate the errors resulting from the triangle inequality above with actual errors observed in simulation showed no correlation at all for a wide range of available molecules.  Further, replacing the bound from the triangle inequality with an alternate estimate using the square-root of the sum of terms continued to show no correlation.  Indeed, we found in the numerical simulations that the error in fact tended to {\it decrease} with larger $N$ for a range of molecules studied.

More precise numerical evidence was obtained from a numerical experiment in which the term order was randomized for the molecule $H_2O$ using a basis with $14$ spin orbitals.  We used the interleaved term order of
Ref.~\onlinecite{HWBT14a}, keeping the ordering of $H_{pp},H_{pqqp},H_{pq},H_{prrq}$ terms fixed, while randomizing the order of the $H_{pqrs}$ terms (randomizing the order of {\it all} terms, not just $H_{pqrs}$ led to a significant increase in numerical error.  We used a second-order TS formula to compare to
Eq.~(\ref{eq:HeffErr}).  From this equaiton, we can see the effect at order $\Delta_t^2$ on the ground state energy of a term
$[[H_\alpha,[H_\beta,H_{\alpha'}]]$ has a random sign depending upon term order.
Thus, this term ordering randomizes the sign of any term involving three distinct $H_{pqrs}$ terms (or involving the commutator of a non-$H_{pqrs}$ term with the commutator of two distinct $H_{pqrs}$ terms).
Further, the signs of distinct terms in Eq.~(\ref{eq:HeffErr}) are decorrelated for each other for most choices:
the average over term orderings of
\be
\bra{\Psi_0}\big[H_\alpha \big[H_\beta,H_{\alpha'}\big] \big]\ket{\Psi_0} \times
\bra{\Psi_0}\big[H_\mu \big[H_\nu,H_{\mu'}\big] \big]\ket{\Psi_0}
\ee
vanishes if $\alpha,\beta,\alpha',\mu,\nu,\nu'$ are all distinct from each other and are all $H_{pqrs}$ terms.

\begin{figure}[t]
\includegraphics[width=8cm]{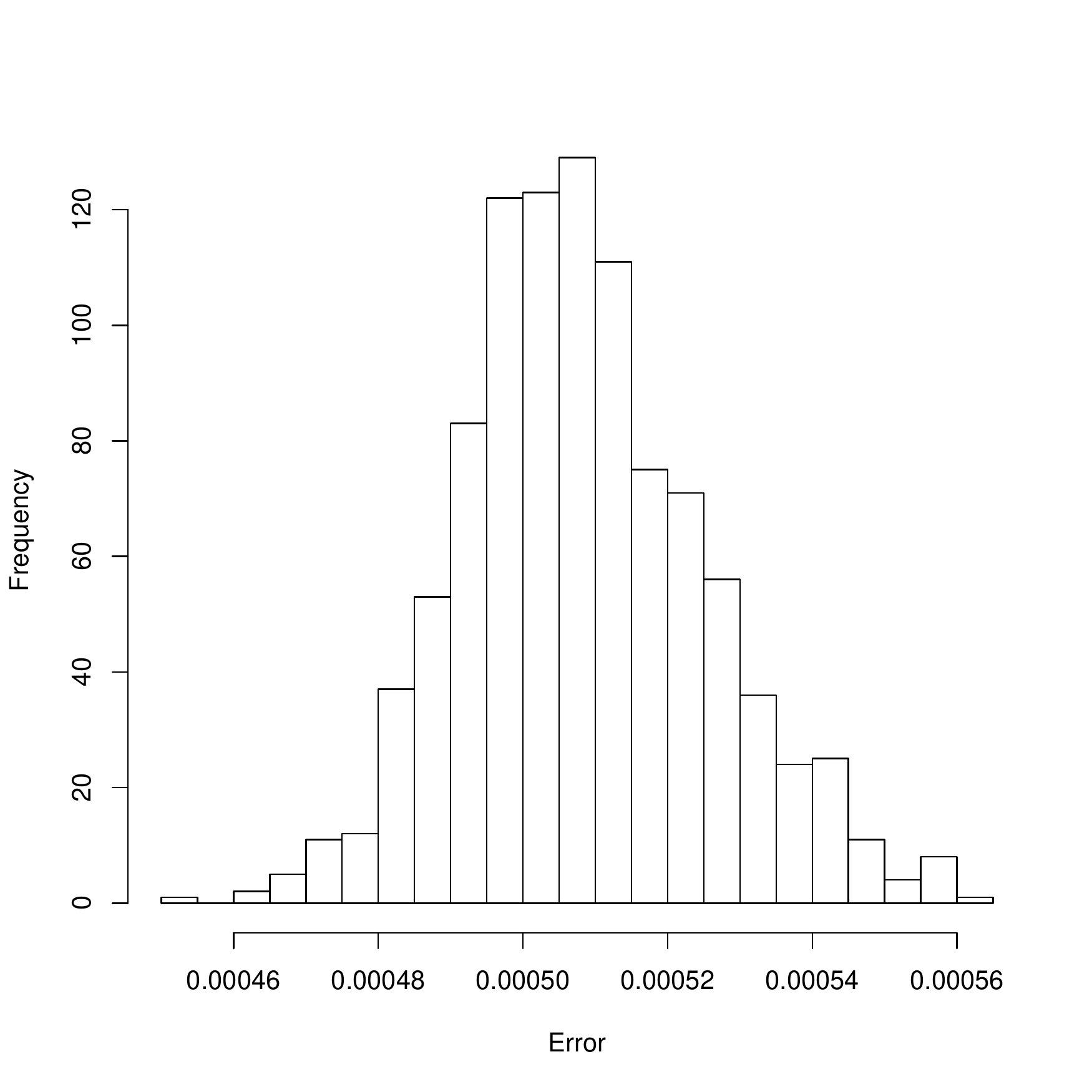}
\caption{Histogram of TS error for $H_2O$ with $H_{pqrs}$ term order randomized, $1000$ samples.}
\label{fig:bell}
\end{figure}

Thus, for a typical term order, we expect the errors to add proportional to $\sqrt{M}$.
Fig.~\ref{fig:bell} shows a histogram of the actual TS error for $1000$ instances.  The Trotter number was set equal to $8$, meaning a time step $\Delta_t=1/8$.  The curve is reasonably close to a Gaussian distribution with a non-zero mean. To quantify the Gaussianity of the curve,
the ratio of the fourth moment to the square of the second moment is equal to $3.15\ldots$, rather than the expected $3$ and the ratio of the third moment to the three-halves power of the second moment is equal $0.32\ldots$ rather than $0$.  The ratio of the root-mean-square width of the curve to the mean is $0.033\ldots$, indicating that the terms with non-zero average still give the dominant contribution to the error at this size.  However, for sufficiently larger sizes, the dominant contribution to the error should indeed arise from double commutators involving three distinct $H_{pqrs}$ terms (as the number of these terms increases rapidly with $N$) and these terms will add with random signs.

It would be interesting to extend this analysis to higher order.  We expect that there will still continue to be an averaging effect.  The order $\Delta_t^4$ correction to the ground state energy is the sum of two terms.  First, there is the ground state expectation value of the order $\Delta_t^4$ correction to Eq.~(\ref{eq:Heff}); this term will still vanish on average over term order.
Second, at order $\Delta_t^4$ there is a correction to the ground state energy which is second order in the order $\Delta_t^2$ Hamiltonian given in Eq.~(\ref{eq:Heff}).  This correction requires summing over intermediate states.  Note, however, that if $\Psi_i$ is an excited state, then
\be
\bra{\Psi_0}\big[H_\alpha \big[H_\beta,H_{\alpha'}\big] \big]\ket{\Psi_i} \times
\bra{\Psi_i}\big[H_\mu \big[H_\nu,H_{\mu'}\big] \big]\ket{\Psi_0}
\ee
vanishes on averaging over term orders if $\alpha,\beta,\alpha',\mu,\nu,\nu'$ are all distinct from each other and are all $H_{pqrs}$ terms.  This holds because,
by the Jacobi identity, the term $[H_\alpha, [H_\beta,H_{\alpha'}] ]$ vanishes identically on averaging over term orders.  However, beyond this
treatment of each term order-by-order, it would be very interesting if an averaging estimate could be given to all orders, similar to the way that
Ref.~\onlinecite{WBCH13a} gave an upper bound in terms of double commutators that was valid to all orders.

As a further test, to see if the bounds were in any way sensitive to molecule geometry or closeness to Hartree-Fock, we studied the molecule ozone, $O_3$.  The kinetics of the recombination of O and O$_2$ to form the O$_3$ ozone molecule depend sensitively on the potential energy surface. In particular the height of a barrier, separating a shallow van der Waals minimum from the ground state, has a big influence on the reaction rate. Accurately calculating the barrier height of this transition state is a challenge for classical calculations, since the full basis is too large to be  treated in a full-configuration interaction calculations and truncated basis sets introduce large approximation errors \cite{Schinke2004}. Calculating the energy of various configuration of the ozone molecule is thus a useful early benchmark problem for a quantum computer. 

For our estimates we  considered three distinct configurations of ozone: a) the ground state , b) the transition state and c) a metastable state at a van der Waals minimum between an $O_2$ molecule and a free oxygen atom. The distances and angles between the atoms at these configurations was obtained from Ref.\cite{Schinke2004}. Direct evaluation of the double commutator bound showed that it was much larger in the ground state than anywhere else. In particular,
the value at the transition state was  7.5 times smaller than in the ground state and at the van der Waals minimum even 9.6 times smaller.
Interestingly, this means that the bound is not in any significant way worse for for the transition point, which is hard to obtain classically.
We used a basis with $60$ spin orbitals for ozone; compared to $H_2O$ in a large basis with $62$ spin orbitals, the bound for the ground state configuration of ozone was roughly twice as large, and it was roughly $0.6$ times as large as that for $Fe_2S_2$ in a basis with $112$ spin orbitals.

\begin{figure}[t]
\includegraphics[width=0.9\linewidth]{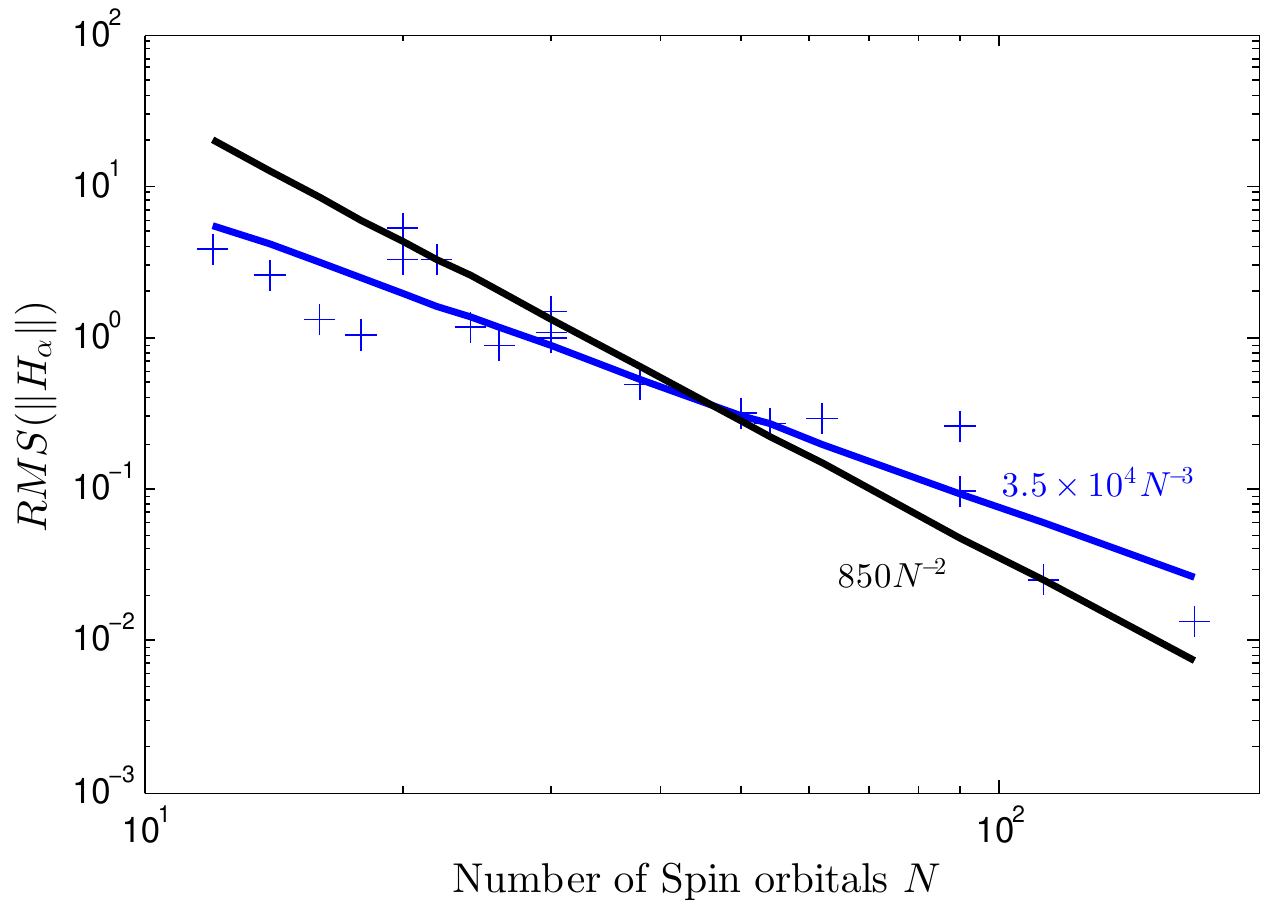}
\caption{RMS value of $\|H_\alpha\|$ for a set of small molecules as a function of the number of spin orbitals $N$.\label{fig:RMS}}
\end{figure}

\subsection{Cauchy--Schwarz bounds and decay of $|h_{pqrs}|^2$}
The Cauchy--Schwarz inequality gives an alternative method for bounding the error in quantum simulations that can be easily computed for large values of $n$.  Let $W(\vec{x})$ be an indicator function that is $1$ if and only if the vector $\vec{x}=(\alpha,\beta,\beta')$ corresponds to a triple of Hamiltonians $H_\alpha, H_\beta, H_{\beta'}$ that contributes to the simulation error in Eq.~(\ref{eq:GSerror1}).
That is, $W(\vec{x})$ is zero if $[[H_{\beta},H_{\alpha}],H_{\beta'}]=0$;
alternately, if we are content to evaluate the error to order $\Delta_t^2$, 
we can set $W(\vec{x})$ to zero if the ground state expectation value of the triple product is known to be zero from symmetry arguments since~\eqref{eq:HeffErr} shows that the ground state energy is unaffected by such terms to order $\Delta_t^3$.  Given these assumptions, the use of the Cauchy--Schwarz inequality applied to triplets which give a non-vanishing contribution to Eq.~(\ref{eq:GSerror1}) shows us that the error can be bounded by the root--mean--square values of $\|H_\alpha\|$ 
\begin{equation}
\|\sum_{\vec{x}}[[H_{\beta},H_{\alpha}],H_{\beta'}]W(\vec{x})\| \le  4\left(\sum_\alpha \|H_\alpha\|^2\right)^{3/2}\!\! \sqrt{N_W},\label{eq:CS}
\end{equation}
where $N_W = \sum_{\vec x} W(\vec x)$.  Similar Cauchy--Schwarz bounds can also be found for Eq.~(\ref{eq:TSerror2}).  This bound has the advantages that it depends on the RMS value of $\|H_\alpha\|$ which is easy to compute directly or from Monte--Carlo sampling and that it also handles constraints in a natural way.

The scaling of the RMS values of $\|H_\alpha\|$ is given in Fig.~\ref{fig:RMS}.  Since the RMS value decreases, the scaling predicted is better than the $\cO(N^{10})$ bound trivially expected.   If we only exclude commuting terms from the sum then $N_W= \cO(N^{10})$.  Using the scalings observed in Fig.~\ref{fig:RMS} and the fact that $H$ consists of $\cO(N^4)$ terms, \eqref{eq:CS} shows that $\delta^{\rm TS}=\cO(N^{\gamma})$, where we find empirically that $2 \leq \gamma \leq 5$, which means the number of Trotter steps needed to achieve a fixed error tolerance is $\cO(N)$-$\cO(N^{2.5})$.  Since $O(N^4)$ gates are needed per TS step, the simulations require a number of gates that is $\cO(N^{5})$-$\cO(N^{6.5})$.  

These scalings also depend strongly on the form of the ground state.  If, for example, we were to assume that only $\cO(N^6)$ terms lead to errors in the ground state energy (which holds when the error in the Hartree--Fock approximation is small) then scaling of the number of gates needed would further drop to $\cO(N^{5.5})$-$\cO(N^{4})$.  Hence properties of the ground state can and should be used to reduce these bounds when possible.

\section{Coalescing}
In Appendix C of Ref.~\onlinecite{WBCH13a}, the idea of ``coalescing" was introduced.  This idea can be regarded as a ``multi-resolution Trotterization":  Rather than trying to determine the minimum TS time step which will work for all terms, we allow different terms to have a different TS step.

One simple realization of the approach within a first-order TS scheme is to pick a fixed time step $\delta_t$ which represents the shortest time that we resolve.  Consider a Hamiltonian $H=\sum_\alpha H_\alpha$.
For each term, $H_\alpha$, we choose some number $n_\alpha$ which reflects how infrequently the term is applied: it will be applied every $n_\alpha$-th step with a strength proportional to $1/n_\alpha$.  Let $\steps$ be the least common multiple of the $n_\alpha$ and let $\Delta_t=\steps \delta_t$.  Then, this scheme gives 
an approximation to evolution over time $\Delta_t$.
For example, if $H=H_1+H_2+H_3$, with $n_1=1,n_2=2,n_3=4$ so that $\Delta_t=4\delta_t$ then we approximate
\begin{eqnarray}
\exp(i 4 \delta_t) & \approx & \Bigl( \exp(i \delta_t H_1) \exp(2 i \delta_t H_2) \exp(4 i \delta_t H_3) \Bigr) \\ \nonumber && \times \Bigl( \exp(i \delta_t H_1) \Bigr) \Bigl( \exp(i \delta_t H_1) \exp(2 i \delta_t H_2) \Bigr) \\ \nonumber && \times \Bigl( \exp(i \delta_t H_1) \Bigr),
\end{eqnarray}
where the parenthesis $\Bigl( \ldots \Bigr)$ are used to separate the four different steps.

Clearly, such a coalescing scheme allows enormous flexibility.  Even within the simple example above, there is room to choose the ordering of terms within each of the four steps (one might choose different orderings in each step).  Further, with a term such as $H_2$, we can choose to execute it on the first and third step as above or on the second and fourth step, and similary we can choose to execute $H_3$ on any of the four steps.

An alternate more complicated coalescing scheme was also presented in Ref.~\onlinecite{WBCH13a}.  One (theoretical) advantage of this more complicated scheme is that it was defined in a way that allowed an inductive proof of tighter upper bounds on the TS error.
For simplicity in this paper, we stick to the simpler first-order approach outlined above.  Also, for simplicity, in all cases we choose the $n_\alpha$ to be powers of two, and we execute a term with given $n_\alpha$ on steps $1, n_\alpha+1, 2n_\alpha+1, \ldots$.  In practice we found no notable advantage to considering other options.
Thus, the important question is how to choose the $n_\alpha$ for a given term.

Using this coalescing scheme, we otherwise continue to followed the interleaved term order of Ref.~\onlinecite{HWBT14a}, so that in a given TS step
we first execute the $H_{pp}$ and $H_{pqqp}$ terms, followed by the $H_{pq}$ terms interleaved with the $H_{prrq}$ terms.  Then, we execute the $H_{pqrs}$ terms.  Coalescing is only applied to the $H_{pqrs}$ terms; that is, all other terms will have $n_\alpha=1$, while for the $H_{pqrs}$ terms we have $n_\alpha=1$ for some terms and $n_\alpha>1$ for others.
We used a first order TS scheme (as noted in Ref.~\onlinecite{HWBT14a} the first order TS offers performance with the same error scaling as second order in this case).

Before going into details, it is worth noting two properties of this scheme.  First, the scheme is exact if all terms $H_\alpha$ commute.  Second, the scheme gives the correct response in the ground state energy to first order for any term $H_\alpha$, regardless of the value of $n_\alpha$.  To make this more precise, suppose that $H_\alpha=\epsilon T$ for some $\alpha$, for operator $T$ and for some $\epsilon << 1$.  Let $\psi_0$ be the approximation to the ground state resulting from TS evolution at $\epsilon=0$.  Then, regardless of $n_\alpha$, the scheme gives a shift in ground state energy equal to $\epsilon \langle \psi_0 | T | \psi_0 \rangle + {\cal O}(\epsilon^2)$.

Finally, one may consider the question of circuits for coalescing.  In Ref.~\onlinecite{HWBT14a}, it was shown that a reduction in circuit depth could be obtained using modified circuits that enable the cancellation of much of the CNOT strings used to perform the Jordan-Wigner transformation and that also enable improved parallelism.  Fortunately, these techniques are compatible with coalescing.  We will find that most of the terms can be aggressively coalesced (choosing a large $n_\alpha)$; for these terms, since many terms will all be executed with the same large $n_\alpha$, most of the CNOT cancellation and parallelization benefits can still be obtained for those terms.  Thus, in what follows, as a proxy for the total depth of the circuit required, we simply use the number of terms in the TS formula, while a more accurate estimate along the lines of previous work such as Ref.~\onlinecite{HWBT14a} would require a detailed analysis of gate depth.

\subsection{Prioritizing Terms}
We now discuss how to choose the $n_\alpha$.  The main improvement here on Ref.~\onlinecite{WBCH13a} is a different way to perform this choice which leads to significant numerical improvements.  Unfortunately, we do not have mathematically rigorous upper bounds to justify our choice; our justification is instead based on extensive numerical simulation for small molecules within reach of a classical simulation.  As explained below, we found a general, fairly simple rule which worked well for all such small molecules, so that we were able to decrease the simulation effort while reducing (or at worst, not increasing) the numerical error in the estimate of the ground state energy.

Previously, in Ref.~\onlinecite{WBCH13a}, it was suggested to coalescing based solely on the {\it magnitude} of the term.  Terms with a larger coefficient would be executed more frequently than those with smaller coefficient.  We chose several different cutoffs $E_1,E_2,E_4,....,E_{\steps}$, and assigned all terms $H_\alpha$ with coefficient greater than or equal to $E_1$ to have $n_\alpha=1$, while all terms $H_\alpha$ with coefficient smaller than $E_1$ but greater than or equal to $E_2$ had $n_2=2$, and so on.  Unfortunately, despite extensive numerical exploration, we were unable to get this scheme to yield any significant improvement.  

In this paper we propose an alternate scheme to choose the $n_\alpha$.  Heuristically, since the scheme gives the correct energy shift to first order, it is important to get the {\it second-order} response to a term correct.  We estimate this effect as follows.  Consider a term $H_\alpha=h_{pqrs} c^{\dagger}_p c^{\dagger}_q c_r c_s$.  We define the importance of the term by a quantity $I_\alpha$ defined as
\be
I_\alpha=\frac{|h_{pqrs}|^2}{\Delta_{p,q,r,s}}
\ee
where $\Delta$ is an estimate of the energy denominator.
We define $\Delta$ as follows.  This is similar to ideas in Section 5 of Ref.\onlinecite{HWBT14a}.
Assume we are working in a Hartree-Fock basis with diagonal terms
\be
\sum_p t_{pp} c^{\dagger}_p c_p + \frac{1}{2}\sum_{p,q} V_{pqqp} c^{\dagger}_p c_p c^{\dagger}_q c_q.
\ee
We let $\omega_p=t_{pp}+\sum_{q \in {\rm occ.}} V_{pqqp}$, where the sum is over occupied orbitals $q$.
We then let
\be
\Delta_{p,q,r,s}=\Bigl| \omega_p+\omega_q-\omega_r-\omega_s \Bigr|.
\ee
This expression describes the second-order response in ground state energy with respect to this perturbation about a Hartree-Fock state.

Our general strategy is to use $I_\alpha$ instead of $h_{pqrs}$.  Thus, even if $h_{pqrs}$ is small, we still regard terms as important if they have a small $\Delta_{p,q,r,s}$, so that terms with a larger $I_\alpha$ are assigned a smaller $n_\alpha$.

One might worry that this approach will not work well if the molecule is far from a Hartree-Fock solution.  However, in strongly interacting Fermi systems, the most important deviations from free fermion behavior arise for states near the Fermi energy.  States sufficiently far above the Fermi energy have occupancy close to zero and those far below the Fermi energy have occupancy close to one, while those near the Fermi energy may have an occupancy far from zero or one in an interacting system.  However, since  this scheme ascribes a large importance to terms involving transitions with all orbitals close to the Fermi energy, we hope that it will continue to work well, although no hard evidence is present.

One important feature of this scheme, as explained below in the section on numerical results, is that we obtain a splitting of the histograms of $I_\alpha$ values.  There are few terms with large $I_\alpha$ (which get $n_\alpha=1$) and many terms with small $I_\alpha$ (which get larger values of $n_\alpha$), with only few terms of intermediate importance.
For these terms of intermediate importance, we have found it useful to define an additional heuristic rule.  While this rule helps, it is not necessary as we have found that fewer terms fall into the intermediate region as molecules grow larger.  This heuristic rule is explaind in the next subsection.

\subsection{Numerical Results}
We now consider several small molecules, and explain specific choices of the cutoffs that lead to improvements using this scheme.

For reasons of the quantum circuits chosen, we make one slight modification to the scheme above.  For us a term in the Hamiltonian
is not simply the term $c^\dagger_p c^\dagger_q c_r c_s+ {\rm h.c.}$ but also includes all other terms involving four fermion operators on spin orbitals $p,q,r,s$ such as $c_p c^\dagger_q c^\dagger_r c_s$.  The reason for this is that some of the same circuits are used to execute both terms.
Hence, we instead choose $\Delta_{p,q,r,s}$ to be the minimum value of $\Delta$ over all such possible assignments of two creation and two annihilation operators to $p,q,r,s$.

Ref.~\onlinecite{whitfield2011} shows that all of the terms for a specific set of $p,q,r,s$ may
be gathered together to form a single circuit (greatly reducing the overall simulation depth).
This leads to a re-write of the original $h_{pqrs}$ values as a set of four strengths
convering the eight basis directions ($xxxx+yyyy$, $xxyy+yyxx$, $yxyx+xyxy$, $yxxy+xyyx$). 
We take the maximum magnitude of these four as representitive of the effect of the combined
terms.

For those terms of intermediate importance, we use the following heuristic rule: a term is chosen to be more important and have $n_\alpha=1$ if it does not annihilates the Hartree-Fock ground state, while it is give a larger value of $n_\alpha$ if it does annihilate the Hartree-Fock ground state.
Whether or not a term annihilate the Hartree-Fock ground state can be determined fairly simply.  For example, if a term $c^\dagger_p c^\dagger_q c_r c_s$ has $p,s$ corresponding to spin up and $q,r$ corresponding to spin down, then it annihilates the Hartree-Fock ground state unless $r,s$ are occupied and $p,q$ are unoccupied.  In fact, since we always ensure that terms are Hermitian and since several different four fermion operators involving spin orbitals $p,q,r,s$ are combined into a single term in the Hamiltonian, such a term will also be retained if one out of $p,q$ is occupied in the Hartree-Fock state and the other is unoccupied and also one out of $r,s$ is occupid and the other is unoccupied.  Similarly, all four spins are up or all four are down, the term is retained if exactly two of the $p,q,r,s$ are occupied in the Hartree-Fock state and the other two are unoccupied.

The final set of rules chosen after some numerical experimentation were: we imposed an upper cutoff $C_U$ and a lower cutoff $C_L$.  For any term with importance  $I_\alpha$ larger than $C_U$, we set $n_\alpha=1$.  For terms with importance in the interval $[C_L,C_U]$, we set $n_\alpha=1$ or $n_\alpha=16$ depending on the heuristic in the above paragraph.  Terms with importance smaller than $C_L$ were always chosen to have $n_\alpha\geq 1$.  Of these terms, we took the $25\%$ with the largest importance and set those to $n_\alpha=16$; of the remaining $75\%$ of the terms, half (or $37.5\%$) were given $n_\alpha=32$ and the remainder were given $n_\alpha=64$.

\begin{figure}[t]
\includegraphics[width=8.5cm]{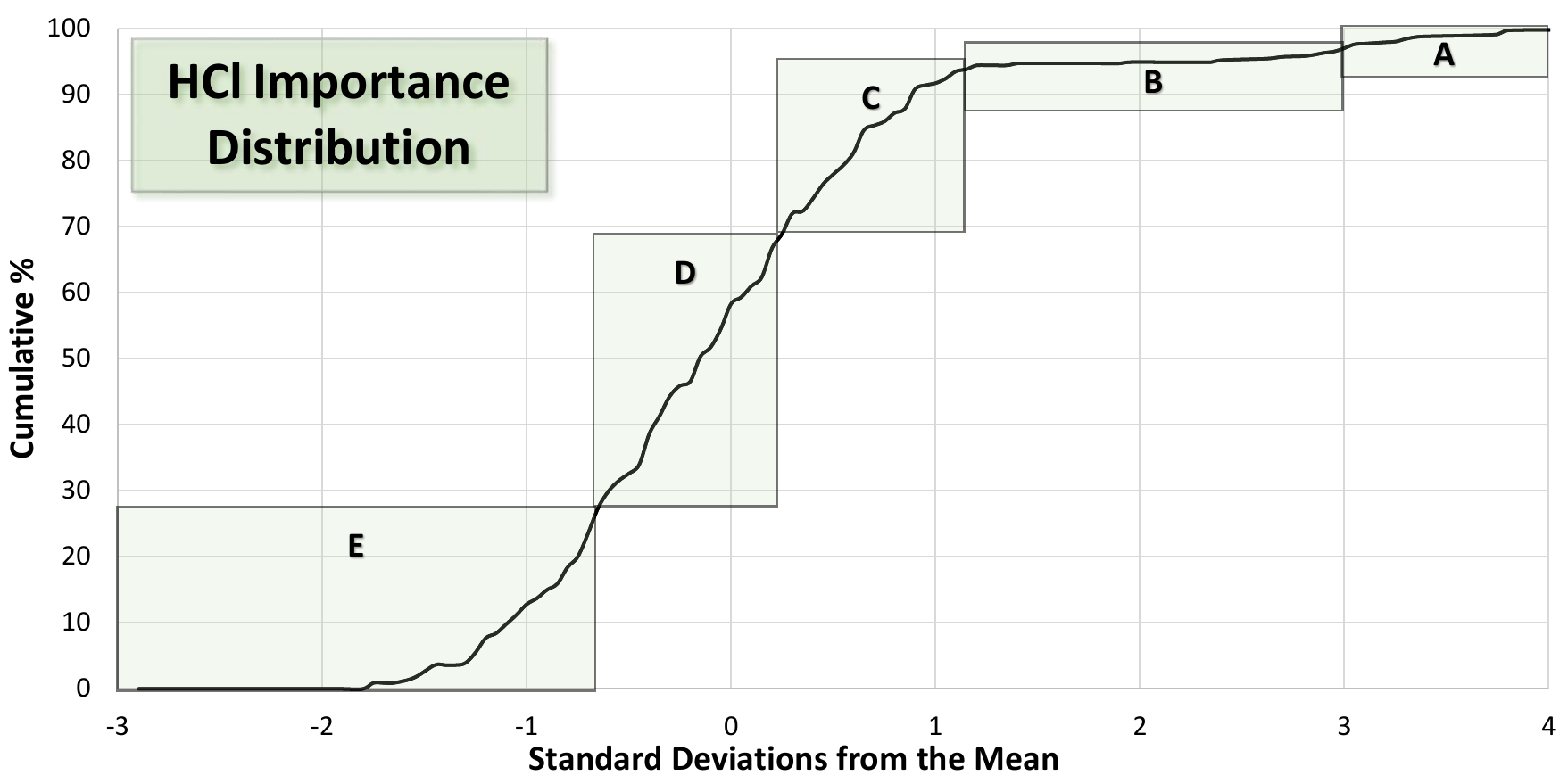}
\caption{Importance Distribution ($h_{pqrs}^2/\omega$) for Hydrogen Chloride. Regions are: (A) terms that must be done every step ($\delta_t$), (B) the ``front porch'' where terms are executed on every step or every 16 steps (based on occupancy), (C) less significant terms (25\% of the remaining) that can be done every 16 steps, (D) $1/2$ of the remaining terms that can be done every 32 steps and (E) all the remaining terms that can be done once every 64 steps.}
\label{figHClPQRS}
\end{figure}

This gives a choice of only 4 possible values of $n_\alpha$: $1,16,32,64$.  A more sophisticated rule with more possible choice might lead to even more speedup.  Our initial studies considered first only two possible values, $1,16$, then later studies considered 3 values, $1,16,32$; in both those cases, the speedup was not as large but still some speedup could be obtained.
The effect of these rules is shown in Fig.~\ref{figHClPQRS} for the molecule $HCl$.

Our goal is to find a choice of $C_U$ and $C_L$ that will reduce the time effort without costing additional accuracy.  Inevitably, the choice of $n_\alpha>1$ for some will reduce the accuracy compared to a choice for $n_\alpha=1$ for all terms, assuming both simulations are run with the same $\delta_t$.  Thus, what we did was to first run a simulation without any coalescing at a fixed value of $\Delta_t$.  Then, we ran a simulation using coalescing, with the above choice of $n_\alpha$, with $\delta_t=\Delta_t/2$.  This permits a more accurate treatement of the terms with highest importance since they use a shorter timestep.
We ran these simulations with $\Delta_t$ sufficiently small that the simulations were in the regime that TS error was proportional to $\Delta_t^2$; once we are in this regime, the relative accuracy of the two simulations (the one with and the one without coalescing) remains unchanged as $\Delta_t$ decreases to this order in $\Delta_t$.

We found a set of rules for $C_U$ and $C_L$ that meant that in every molecule we tried, the error did not increase using coalescing, and in many cases it decreased.  We chose $\log(C_U)$ equal to the average value of the log of $I_\alpha$ plus three times the standard deviation of the log of $I_\alpha$, and we chose $\log(C_L)$ equal to the average value of the log of $I_\alpha$ plus $1.2$ times the standard deviation.
Specific molecules simulated using \lqd were 
$HF, H_2O, NH_3, NCl, F_2$, and $H_2S$.

\begin{figure}[t]
\includegraphics[width=9cm]{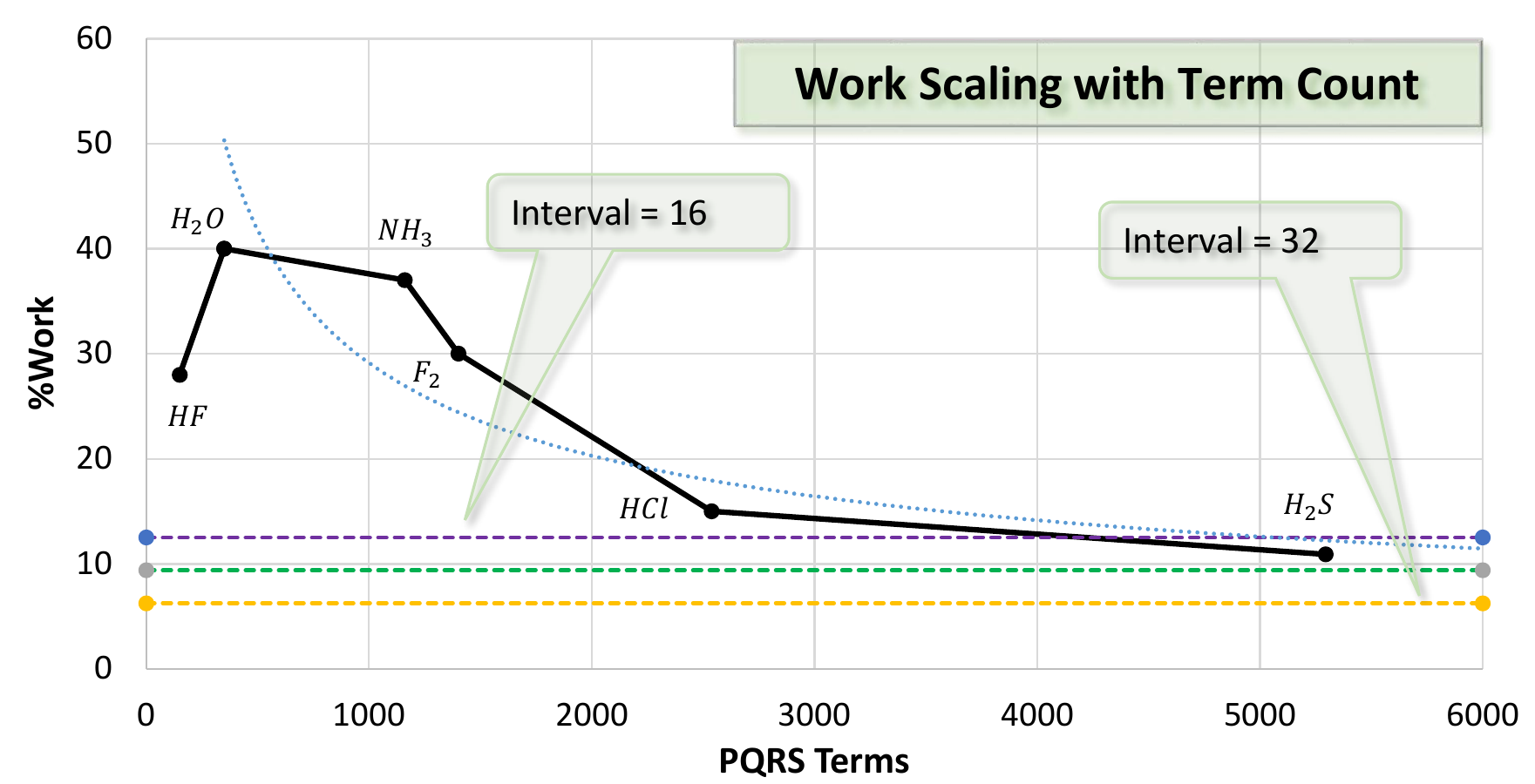}
\caption{Amount of work for various molecules using the current scheme at a Trotter number of 64. The three horizontal dotted lines are asymptotes for doing all terms at an interval of 16 steps (top) and 32 steps (bottom). The current approach (shown in the previous figure) is close to the middle asymptote for a 50/50 mixture of the two limits.}
\label{figWork}
\end{figure}

The result for the work required is shown in Fig.~\ref{figWork}.
By hand-tuning the choice of $C_U,C_L$ for specific molecules it is possible to further reduce the work without increasing the error.  However, this is clearly an unrealistic test: since our goal is to develop rules that will be useful to reduce work on a real quantum computer, there is no way to know in advance what the most optimal values are.
However, this general rule works well for all these molecules and is close to optimal.

\begin{figure}[t]
\includegraphics[width=8.5cm]{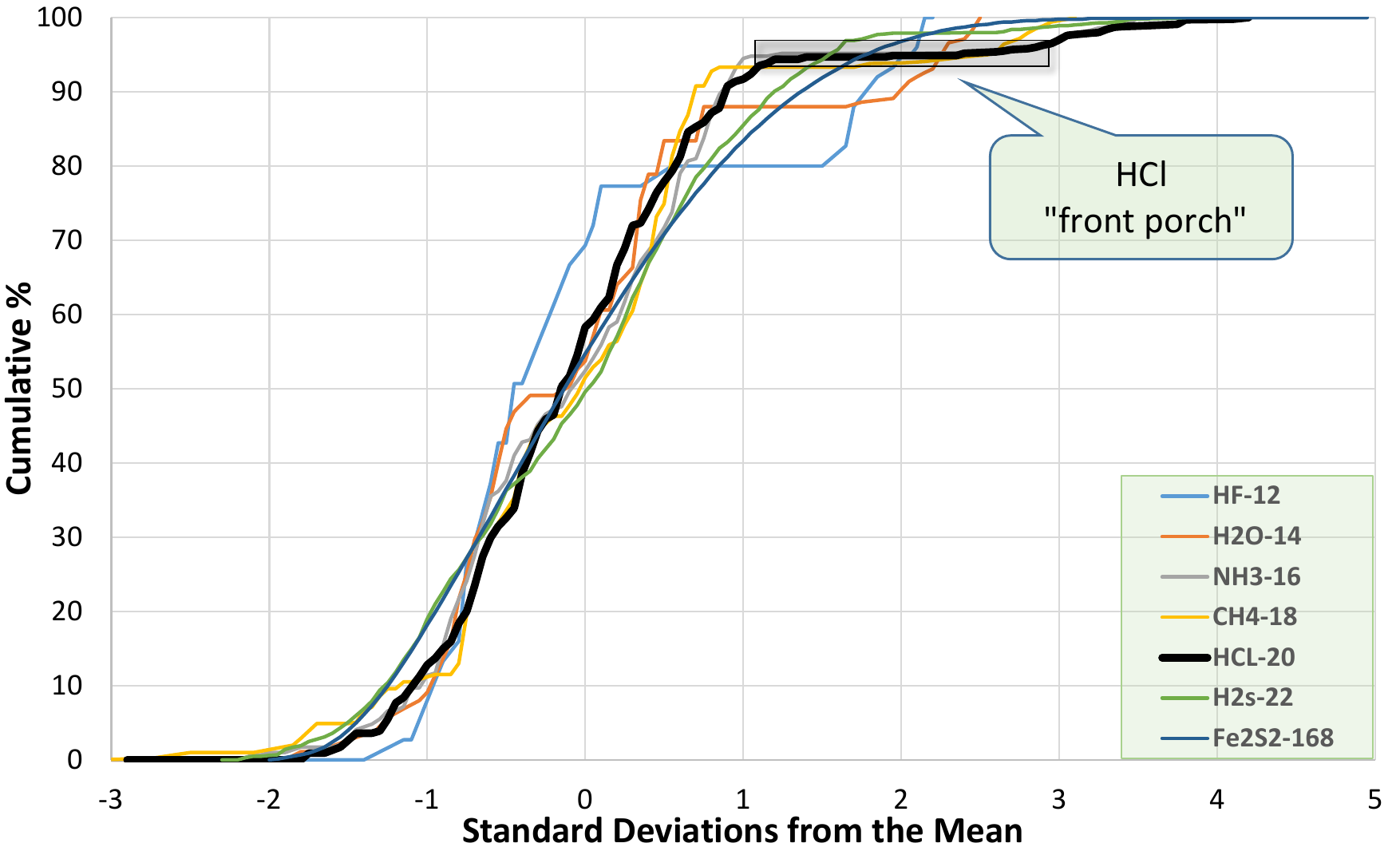}
\caption{Results for several molecules of different sizes. Note that the ``front porch'' disappears as $N$ increases and may thus be ignored as we scale (terms are either coalesced or not, there is no need for occupancy calculations).  Further, the most important terms become more important relative to the mean (and hence the typical terms are less important relative to the most important ones) as $N$ increases.}
\label{manyMolecules}
\end{figure}

We cannot simulate larger molecules, but assuming that the scheme does continue to work, we can investigate what speedups would be achieved.
The first important point is that while the mean of $\log(I_\alpha)$ is observed to decrease with increasing $N$, no clear trend was observed for the standard deviation.  Instead, the standard deviation was observed to vary between roughly $2-3$ on a log-base $10$ scale with no clear trend, considering larger molecules up to $Fe_2S_2$ in a basis with $168$ spin-ortbitals.
After normalizing $\log(I_\alpha)$ by subtracting the mean value for the given molecule and dividing by the standard deviation (so that we use $(\log(I_\alpha)-\overline{\log(I_\alpha)})/{\rm var}(\log(I_\alpha))$ as the horizontal axis), the results are as shown in Fig.~\ref{manyMolecules}.
One observes a range of importance (appearing as a flat spot in the curve, which we term the ``front porch" in the figures) into which few terms fall.  This range is observed to move to larger importance relative to the mean as the molecue size increases.  The rules for $C_U,C_L$ above were chosen such that for the smaller molecules, the interval $[C_L,C_U]$ is roughly the region of this front porch.  Since the width of the front porch decreases, relatively fewer terms fall into this region.
Further, as molecule size increases, one observes that the importance (relative to the mean) of the most importand terms grows larger (see $Fe_2S_2$ for example, which extends furthest to the right on the curve).  Since then these large molecules have a few terms with very high importance, this suggests that for these molecules it will be possible to coalesce almost all terms except these few high importance terms leading to potentially even larger gains in runtime for larger molecules.

\section{Discussion and conclusion}

The Hamiltonian of a small molecule contains a number of terms scaling as $N^4$ with the number of orbitals $N$ in the simulation, which represents a major bottleneck to quantum simulations. Previous analysis \cite{WBCH13a,HWBT14a} predict a general complexity in gate count (which can be parallelized saving a factor of $N$) scaling as $N^8$-$N^9$. By numerically evaluating an upper bound on the simulation error entailed by the TS approximation, we have demonstrated that the scaling is in fact much more favorable for real-world molecules, in the $N^{5.5}$-$N^{6.5}$ range in worst case. Evaluation of this bound on a greater number of molecules would be necessary to reach firm conclusions; further, a better understanding would be needed of whether the error terms should be added in absolute value or whether they add with random signs. 

Because the scaling analysis of \cite{WBCH13a} uses molecules drawn from a random ensemble, our observations strongly indicates that this ensemble fails to accurately reproduce the statistical properties of real molecules. As a simple indication of this discrepancy, we find that the sum of the magnitude of all the Hamiltonian coefficients $\sum_{pqrs} |h_{pqrs}|$ scales like $N^2$ for real molecules, while the random ensemble yields $N^4$ by design. 

We have explored the consequences of breaking up the Hamiltonian into a different sum of terms to implement the TS decomposition. With artificial molecules and a few real molecules, we have seen that this alternative decomposition can offer significant savings. Thus, we believe that exploring the different ways in which this decomposition can be realized is a good approach to obtain further improvements. This decomposition is usually guided by our ability to simulate sparse Hamiltonians \cite{AT03b}. It is noteworthy that in the decomposition we use, each term $G_\alpha^2$ is no sparser than the full Hamiltonian $H$. This illustrates that other criteria should be envisioned to guide this decomposition.

We have also developed the coalescing technique, showing large gains for small molecules.  It seems likely based on our numerical data that even more aggressive coalescing will be possible on larger molecules, leading to further gains.  We have separately explored the possibility of combining coalescing with the sum of squares decomposition of the Hamiltonian, but have not found any improvement this way.  Combining the coalescing technique with the improved scaling here suggests that simulation of large molecules with the order of a hundreds of spin orbitals will be much more practical than indicated in \cite{WBCH13a}.

Lastly, our study, as well as previous ones \cite{ADLH05a,WBA10a,KWPY11a,WBCH13a,HWBT14a}, have focused on low-order TS decompositions. While schemes based on random walks \cite{BC12a} or techniques from simulating continuous query algorithms \cite{BCCK13a} promise lower query complexity than TS based simulations for general purpose quantum simulations, their reliance on a quantum oracle makes a comparison of their time complexity to that of TS algorithms for quantum chemistry challenging. Their concrete realization would therefore require a circuit which, given inputs $(i,j)$, returns the matrix element $\bra i H \ket j$. Since an arbitrary Hamiltonian may contain $\sim N^4$ non-zero terms, such a circuit would at best require $\sim N^4$ gates, yielding an overall complexity greater than what we have observed here.  This can be parallelized using, for example, a QRAM\cite{GLM08}, but at the cost of an enormous space overhead.  Despite these difficulties in implementing the oracles, it remains possible that these simulation techniques could be competitive because they may perform better than what their upper bounds suggest.

{\it Acknowledgement---}
DP acknowledges the hospitality of the University of Sydney where this research project was realized. DP and ACD were supported by the ARC via the Centre of Excellence in Engineered Quantum Systems (EQuS), project number CE110001013.

\appendix
\section{Estimate for Norm of Random Hamiltonian}
We now briefly discuss the extent to which we can prove $\Vert H_{>\alpha} \Vert \leq \cO(N^2)$.  Of course, physically we expect such a bound to hold, especially since the electron-electron interaction has norm $\cO(N^2)$.  However, it is interesting to consider the extent to which we can show it for a random ensemble.

Proving this bound $\cO(N^2)$ might be difficult but it is possible using a version of the trace method to prove the weaker bound $\Vert H_{>\alpha} \Vert \leq \cO(N^{5/2})$ using a version of the trace method.  Consider the average (over random choices of the Hamiltonian) of ${\rm tr}(H_{>\alpha}^n)$ for a constant $n$ chosen later.  Expanding the trace as a sum
$\sum_{\alpha_1>\alpha} ... \sum_{\alpha_n>\alpha} {\rm tr}(H_{\alpha_1} ... H_{\alpha_n})$, the only terms that do not vanish on average are where the sequence $\alpha_1,...,\alpha_n$ repeats each $\alpha$ an {\it even} number of times.  Hence, of the at most $m^n$ terms in the sum, only at most $(mn)^{n/2}$ are non-vanishing.  Let us use an overline to denote the average over Hamiltonians.  Thus $\overline{{\rm tr}(H_{>\alpha}^n)}$ is bounded by $2^N \cdot ({\rm const.} \times mn)^{n/2}$ and  $\Bigl(\overline{{\rm tr}(H_{>\alpha}^n)} \Bigr)^{1/n} \leq {\rm const.} \times 2^{N/n} (mn)^{1/2}$.
We will choose $n=2N \ln(2)$ so that
$\Bigl(\overline{{\rm tr}(H_{>\alpha}^n)} \Bigr)^{1/n} \leq \cO(N^{5/2})$ and so $\overline{\Vert H_{>\alpha} \Vert} \leq \cO(N^{5/2})$.
Further, using similar estimates, one can show that with high probability $\Vert H_{>\alpha} \Vert=\cO(N^{5/2})$; to see this, by Markov's inequality, the probability that $\Vert H_{>\alpha} \Vert$ is, for example, twice as big as $\Bigl(\overline{{\rm tr}(H_{>\alpha}^n)} \Bigr)^{1/n}$ is at most $2^{-n}$.

Even using this weaker bound $\Vert H_{>\alpha} \Vert \leq {\cal O}(N^{5/2})$ we still find the bound $\sum_{\alpha} \Vert [[H_{>\alpha},H_\alpha],H_{>\alpha}]] \Vert \leq {\cal O}(N^9)$ which still improves on the triangle inequality estimate $\cO(N^{10})$.

\narrowtext
\bibliographystyle{apsrev4-1}
\bibliography{qubib}

\begin{thebibliography}{24}%
\makeatletter
\providecommand \@ifxundefined [1]{%
 \@ifx{#1\undefined}
}%
\providecommand \@ifnum [1]{%
 \ifnum #1\expandafter \@firstoftwo
 \else \expandafter \@secondoftwo
 \fi
}%
\providecommand \@ifx [1]{%
 \ifx #1\expandafter \@firstoftwo
 \else \expandafter \@secondoftwo
 \fi
}%
\providecommand \natexlab [1]{#1}%
\providecommand \enquote  [1]{``#1''}%
\providecommand \bibnamefont  [1]{#1}%
\providecommand \bibfnamefont [1]{#1}%
\providecommand \citenamefont [1]{#1}%
\providecommand \href@noop [0]{\@secondoftwo}%
\providecommand \href [0]{\begingroup \@sanitize@url \@href}%
\providecommand \@href[1]{\@@startlink{#1}\@@href}%
\providecommand \@@href[1]{\endgroup#1\@@endlink}%
\providecommand \@sanitize@url [0]{\catcode `\\12\catcode `\$12\catcode
  `\&12\catcode `\#12\catcode `\^12\catcode `\_12\catcode `\%12\relax}%
\providecommand \@@startlink[1]{}%
\providecommand \@@endlink[0]{}%
\providecommand \url  [0]{\begingroup\@sanitize@url \@url }%
\providecommand \@url [1]{\endgroup\@href {#1}{\urlprefix }}%
\providecommand \urlprefix  [0]{URL }%
\providecommand \Eprint [0]{\href }%
\providecommand \doibase [0]{http://dx.doi.org/}%
\providecommand \selectlanguage [0]{\@gobble}%
\providecommand \bibinfo  [0]{\@secondoftwo}%
\providecommand \bibfield  [0]{\@secondoftwo}%
\providecommand \translation [1]{[#1]}%
\providecommand \BibitemOpen [0]{}%
\providecommand \bibitemStop [0]{}%
\providecommand \bibitemNoStop [0]{.\EOS\space}%
\providecommand \EOS [0]{\spacefactor3000\relax}%
\providecommand \BibitemShut  [1]{\csname bibitem#1\endcsname}%
\let\auto@bib@innerbib\@empty
\bibitem [{\citenamefont {Wecker}\ \emph {et~al.}(2013)\citenamefont {Wecker},
  \citenamefont {Bauer}, \citenamefont {Clark}, \citenamefont {Hastings},\ and\
  \citenamefont {Troyer}}]{WBCH13a}%
  \BibitemOpen
  \bibfield  {author} {\bibinfo {author} {\bibfnamefont {D.}~\bibnamefont
  {Wecker}}, \bibinfo {author} {\bibfnamefont {B.}~\bibnamefont {Bauer}},
  \bibinfo {author} {\bibfnamefont {B.~K.}\ \bibnamefont {Clark}}, \bibinfo
  {author} {\bibfnamefont {M.~B.}\ \bibnamefont {Hastings}}, \ and\ \bibinfo
  {author} {\bibfnamefont {M.}~\bibnamefont {Troyer}},\ }\href
  {http://arxiv.org/abs/1312.1695} {\enquote {\bibinfo {title} {Can quantum
  chemistry be performed on a small quantum computer?}}\ } (\bibinfo {year}
  {2013}),\ \Eprint {http://arxiv.org/abs/arXiv:1312.1695} {arXiv:1312.1695}
  \BibitemShut {NoStop}%
\bibitem [{\citenamefont {Hastings}\ \emph {et~al.}(2014)\citenamefont
  {Hastings}, \citenamefont {Wecker}, \citenamefont {Bauer},\ and\
  \citenamefont {Troyer}}]{HWBT14a}%
  \BibitemOpen
  \bibfield  {author} {\bibinfo {author} {\bibfnamefont {M.~B.}\ \bibnamefont
  {Hastings}}, \bibinfo {author} {\bibfnamefont {D.}~\bibnamefont {Wecker}},
  \bibinfo {author} {\bibfnamefont {B.}~\bibnamefont {Bauer}}, \ and\ \bibinfo
  {author} {\bibfnamefont {M.}~\bibnamefont {Troyer}},\ }\href@noop {}
  {\enquote {\bibinfo {title} {Improving quantum algorithms for quantum
  chemistry},}\ } (\bibinfo {year} {2014}),\ \Eprint
  {http://arxiv.org/abs/arXiv:1403.1539} {arXiv:1403.1539} \BibitemShut
  {NoStop}%
\bibitem [{\citenamefont {Feynman}(1982)}]{Fey82a}%
  \BibitemOpen
  \bibfield  {author} {\bibinfo {author} {\bibfnamefont {R.~P.}\ \bibnamefont
  {Feynman}},\ }\href@noop {} {\bibfield  {journal} {\bibinfo  {journal} {Int.
  J. of Theor. Phys.}\ }\textbf {\bibinfo {volume} {21}},\ \bibinfo {pages}
  {467} (\bibinfo {year} {1982})}\BibitemShut {NoStop}%
\bibitem [{\citenamefont {Lloyd}(1996)}]{Llo96b}%
  \BibitemOpen
  \bibfield  {author} {\bibinfo {author} {\bibfnamefont {S.}~\bibnamefont
  {Lloyd}},\ }\href@noop {} {\bibfield  {journal} {\bibinfo  {journal}
  {Science}\ }\textbf {\bibinfo {volume} {273}},\ \bibinfo {pages} {1073}
  (\bibinfo {year} {1996})}\BibitemShut {NoStop}%
\bibitem [{\citenamefont {Zalka}(1998)}]{Zal98a}%
  \BibitemOpen
  \bibfield  {author} {\bibinfo {author} {\bibfnamefont {C.}~\bibnamefont
  {Zalka}},\ }\href@noop {} {\bibfield  {journal} {\bibinfo  {journal}
  {Fortsch. Phys.}\ }\textbf {\bibinfo {volume} {46}},\ \bibinfo {pages} {877}
  (\bibinfo {year} {1998})}\BibitemShut {NoStop}%
\bibitem [{\citenamefont {Ortiz}\ \emph {et~al.}(2001)\citenamefont {Ortiz},
  \citenamefont {Gubernatis}, \citenamefont {Knill},\ and\ \citenamefont
  {Laflamme}}]{OGK+01a}%
  \BibitemOpen
  \bibfield  {author} {\bibinfo {author} {\bibfnamefont {G.}~\bibnamefont
  {Ortiz}}, \bibinfo {author} {\bibfnamefont {J.~E.}\ \bibnamefont
  {Gubernatis}}, \bibinfo {author} {\bibfnamefont {E.}~\bibnamefont {Knill}}, \
  and\ \bibinfo {author} {\bibfnamefont {R.}~\bibnamefont {Laflamme}},\
  }\href@noop {} {\bibfield  {journal} {\bibinfo  {journal} {Phys. Rev. A}\
  }\textbf {\bibinfo {volume} {64}},\ \bibinfo {pages} {22319} (\bibinfo {year}
  {2001})},\ \Eprint {http://arxiv.org/abs/cond-mat/0012334} {cond-mat/0012334}
  \BibitemShut {NoStop}%
\bibitem [{\citenamefont {Aharonov}\ and\ \citenamefont
  {Ta-Shma}(2003)}]{AT03b}%
  \BibitemOpen
  \bibfield  {author} {\bibinfo {author} {\bibfnamefont {D.}~\bibnamefont
  {Aharonov}}\ and\ \bibinfo {author} {\bibfnamefont {A.}~\bibnamefont
  {Ta-Shma}},\ }\href@noop {} {\bibfield  {journal} {\bibinfo  {journal} {Proc.
  35th Annual ACM Symp. on Theo. Comp.}\ ,\ \bibinfo {pages} {20}} (\bibinfo
  {year} {2003})}\BibitemShut {NoStop}%
\bibitem [{\citenamefont {Wiebe}\ \emph {et~al.}(2011)\citenamefont {Wiebe},
  \citenamefont {Berry}, \citenamefont {H{\o}yer},\ and\ \citenamefont
  {Sanders}}]{WBHS11a}%
  \BibitemOpen
  \bibfield  {author} {\bibinfo {author} {\bibfnamefont {N.}~\bibnamefont
  {Wiebe}}, \bibinfo {author} {\bibfnamefont {D.~W.}\ \bibnamefont {Berry}},
  \bibinfo {author} {\bibfnamefont {P.}~\bibnamefont {H{\o}yer}}, \ and\
  \bibinfo {author} {\bibfnamefont {B.~C.}\ \bibnamefont {Sanders}},\
  }\href@noop {} {\bibfield  {journal} {\bibinfo  {journal} {J. Phys. A}\
  }\textbf {\bibinfo {volume} {44}},\ \bibinfo {pages} {445308} (\bibinfo
  {year} {2011})}\BibitemShut {NoStop}%
\bibitem [{\citenamefont {Berry}\ and\ \citenamefont {Childs}(2012)}]{BC12a}%
  \BibitemOpen
  \bibfield  {author} {\bibinfo {author} {\bibfnamefont {D.}~\bibnamefont
  {Berry}}\ and\ \bibinfo {author} {\bibfnamefont {A.}~\bibnamefont {Childs}},\
  }\href@noop {} {\bibfield  {journal} {\bibinfo  {journal} {Quant. Info. and
  Comp.}\ }\textbf {\bibinfo {volume} {12}},\ \bibinfo {pages} {29} (\bibinfo
  {year} {2012})}\BibitemShut {NoStop}%
\bibitem [{\citenamefont {Berry}\ \emph {et~al.}(2013)\citenamefont {Berry},
  \citenamefont {Childs}, \citenamefont {Cleve}, \citenamefont {Kothari},\ and\
  \citenamefont {Somma}}]{BCCK13a}%
  \BibitemOpen
  \bibfield  {author} {\bibinfo {author} {\bibfnamefont {D.~W.}\ \bibnamefont
  {Berry}}, \bibinfo {author} {\bibfnamefont {A.~M.}\ \bibnamefont {Childs}},
  \bibinfo {author} {\bibfnamefont {R.}~\bibnamefont {Cleve}}, \bibinfo
  {author} {\bibfnamefont {R.}~\bibnamefont {Kothari}}, \ and\ \bibinfo
  {author} {\bibfnamefont {R.~D.}\ \bibnamefont {Somma}},\ }\href@noop {}
  {\enquote {\bibinfo {title} {Exponential improvement in precision for
  simulating sparse hamiltonians},}\ } (\bibinfo {year} {2013}),\ \Eprint
  {http://arxiv.org/abs/arXiv:1312.1414} {arXiv:1312.1414} \BibitemShut
  {NoStop}%
\bibitem [{\citenamefont {Aspuru-Guzik}\ \emph {et~al.}(2005)\citenamefont
  {Aspuru-Guzik}, \citenamefont {Dutoi}, \citenamefont {Love},\ and\
  \citenamefont {Head-Gordon}}]{ADLH05a}%
  \BibitemOpen
  \bibfield  {author} {\bibinfo {author} {\bibfnamefont {A.}~\bibnamefont
  {Aspuru-Guzik}}, \bibinfo {author} {\bibfnamefont {A.~D.}\ \bibnamefont
  {Dutoi}}, \bibinfo {author} {\bibfnamefont {P.~J.}\ \bibnamefont {Love}}, \
  and\ \bibinfo {author} {\bibfnamefont {M.}~\bibnamefont {Head-Gordon}},\
  }\href@noop {} {\bibfield  {journal} {\bibinfo  {journal} {Science}\ }\textbf
  {\bibinfo {volume} {309}},\ \bibinfo {pages} {1704} (\bibinfo {year}
  {2005})}\BibitemShut {NoStop}%
\bibitem [{\citenamefont {Whitfield}\ \emph {et~al.}(2010)\citenamefont
  {Whitfield}, \citenamefont {Biamonte},\ and\ \citenamefont
  {Aspuru-Guzik}}]{WBA10a}%
  \BibitemOpen
  \bibfield  {author} {\bibinfo {author} {\bibfnamefont {J.}~\bibnamefont
  {Whitfield}}, \bibinfo {author} {\bibfnamefont {J.}~\bibnamefont {Biamonte}},
  \ and\ \bibinfo {author} {\bibfnamefont {A.}~\bibnamefont {Aspuru-Guzik}},\
  }\href@noop {} {\bibfield  {journal} {\bibinfo  {journal} {Molecular Phys.}\
  }\textbf {\bibinfo {volume} {109}},\ \bibinfo {pages} {735} (\bibinfo {year}
  {2010})},\ \Eprint {http://arxiv.org/abs/arXiv:1001.3855} {arXiv:1001.3855}
  \BibitemShut {NoStop}%
\bibitem [{\citenamefont {Kassal}\ \emph {et~al.}(2011)\citenamefont {Kassal},
  \citenamefont {Whitfield}, \citenamefont {Perdomo-Ortiz}, \citenamefont
  {Yung},\ and\ \citenamefont {Aspuru-Guzik}}]{KWPY11a}%
  \BibitemOpen
  \bibfield  {author} {\bibinfo {author} {\bibfnamefont {I.}~\bibnamefont
  {Kassal}}, \bibinfo {author} {\bibfnamefont {J.~D.}\ \bibnamefont
  {Whitfield}}, \bibinfo {author} {\bibfnamefont {A.}~\bibnamefont
  {Perdomo-Ortiz}}, \bibinfo {author} {\bibfnamefont {M.-H.}\ \bibnamefont
  {Yung}}, \ and\ \bibinfo {author} {\bibfnamefont {A.}~\bibnamefont
  {Aspuru-Guzik}},\ }\href@noop {} {\bibfield  {journal} {\bibinfo  {journal}
  {Annual Review of Physical Chemistry}\ }\textbf {\bibinfo {volume} {62}},\
  \bibinfo {pages} {185} (\bibinfo {year} {2011})}\BibitemShut {NoStop}%
\bibitem [{\citenamefont {Gan}\ and\ \citenamefont {Harrison}(2005)}]{GH05a}%
  \BibitemOpen
  \bibfield  {author} {\bibinfo {author} {\bibfnamefont {Z.}~\bibnamefont
  {Gan}}\ and\ \bibinfo {author} {\bibfnamefont {R.~J.}\ \bibnamefont
  {Harrison}},\ }in\ \href@noop {} {\emph {\bibinfo {booktitle} {Proc.
  {ACM}/{IEEE} {SC} 2005 Conference}}}\ (\bibinfo {year} {2005})\ p.~\bibinfo
  {pages} {22}\BibitemShut {NoStop}%
\bibitem [{\citenamefont {Nakano}\ and\ \citenamefont {Sakai}(2011)}]{NS11a}%
  \BibitemOpen
  \bibfield  {author} {\bibinfo {author} {\bibfnamefont {H.}~\bibnamefont
  {Nakano}}\ and\ \bibinfo {author} {\bibfnamefont {T.}~\bibnamefont {Sakai}},\
  }\href@noop {} {\bibfield  {journal} {\bibinfo  {journal} {J. Phys. Soc.
  Japan}\ }\textbf {\bibinfo {volume} {80}},\ \bibinfo {pages} {053704}
  (\bibinfo {year} {2011})}\BibitemShut {NoStop}%
\bibitem [{\citenamefont {L{\"a}uchli}\ \emph {et~al.}(2011)\citenamefont
  {L{\"a}uchli}, \citenamefont {Sudan},\ and\ \citenamefont
  {S{\o}rensen}}]{LSS11a}%
  \BibitemOpen
  \bibfield  {author} {\bibinfo {author} {\bibfnamefont {A.~M.}\ \bibnamefont
  {L{\"a}uchli}}, \bibinfo {author} {\bibfnamefont {J.}~\bibnamefont {Sudan}},
  \ and\ \bibinfo {author} {\bibfnamefont {E.~S.}\ \bibnamefont
  {S{\o}rensen}},\ }\href@noop {} {\bibfield  {journal} {\bibinfo  {journal}
  {Phys. Rev. B}\ }\textbf {\bibinfo {volume} {83}},\ \bibinfo {pages} {212401}
  (\bibinfo {year} {2011})}\BibitemShut {NoStop}%
\bibitem [{\citenamefont {Capponi}\ \emph {et~al.}(2013)\citenamefont
  {Capponi}, \citenamefont {Derzhko}, \citenamefont {Honecker}, \citenamefont
  {L{\"a}uchli},\ and\ \citenamefont {Richter}}]{CDHL13a}%
  \BibitemOpen
  \bibfield  {author} {\bibinfo {author} {\bibfnamefont {S.}~\bibnamefont
  {Capponi}}, \bibinfo {author} {\bibfnamefont {O.}~\bibnamefont {Derzhko}},
  \bibinfo {author} {\bibfnamefont {A.}~\bibnamefont {Honecker}}, \bibinfo
  {author} {\bibfnamefont {A.~M.}\ \bibnamefont {L{\"a}uchli}}, \ and\ \bibinfo
  {author} {\bibfnamefont {J.}~\bibnamefont {Richter}},\ }\href@noop {}
  {\bibfield  {journal} {\bibinfo  {journal} {Phys. Rev. B}\ }\textbf {\bibinfo
  {volume} {88}},\ \bibinfo {pages} {144416} (\bibinfo {year}
  {2013})}\BibitemShut {NoStop}%
\bibitem [{\citenamefont {Kurashige}\ \emph {et~al.}(2013)\citenamefont
  {Kurashige}, \citenamefont {Chan},\ and\ \citenamefont {Yanai}}]{KCY13a}%
  \BibitemOpen
  \bibfield  {author} {\bibinfo {author} {\bibfnamefont {Y.}~\bibnamefont
  {Kurashige}}, \bibinfo {author} {\bibfnamefont {G.~K.-L.}\ \bibnamefont
  {Chan}}, \ and\ \bibinfo {author} {\bibfnamefont {T.}~\bibnamefont {Yanai}},\
  }\href@noop {} {\bibfield  {journal} {\bibinfo  {journal} {Nature Chemistry}\
  }\textbf {\bibinfo {volume} {5}},\ \bibinfo {pages} {660} (\bibinfo {year}
  {2013})}\BibitemShut {NoStop}%
\bibitem [{Note1()}]{Note1}%
  \BibitemOpen
  \bibinfo {note} {This constant average cost can only be achieved when a
  sequence of free evolution operators are executed, which will always be the
  case here. Otherwise the cost is $\sim N$ due to the gates that are required
  to implement the Jordan-Wigner mapping of fermion orbitals into
  qubits.}\BibitemShut {Stop}%
\bibitem [{\citenamefont {Horn}\ and\ \citenamefont {Johnson}(1985)}]{HJ85a}%
  \BibitemOpen
  \bibfield  {author} {\bibinfo {author} {\bibfnamefont {R.}~\bibnamefont
  {Horn}}\ and\ \bibinfo {author} {\bibfnamefont {C.}~\bibnamefont {Johnson}},\
  }\href@noop {} {\emph {\bibinfo {title} {Matrix Analysis}}}\ (\bibinfo
  {publisher} {Cambridge University Press, Cambridge},\ \bibinfo {year}
  {1985})\BibitemShut {NoStop}%
\bibitem [{\citenamefont {Wecker}\ and\ \citenamefont {Svore}()}]{liquidref}%
  \BibitemOpen
  \bibfield  {author} {\bibinfo {author} {\bibfnamefont {D.}~\bibnamefont
  {Wecker}}\ and\ \bibinfo {author} {\bibfnamefont {K.~M.}\ \bibnamefont
  {Svore}},\ }\href@noop {} {\ }\Eprint {http://arxiv.org/abs/1402.4467}
  {arXiv:1402.4467} \BibitemShut {NoStop}%
\bibitem [{\citenamefont {Schinke}\ and\ \citenamefont
  {Fleurat-Lessard}(2004)}]{Schinke2004}%
  \BibitemOpen
  \bibfield  {author} {\bibinfo {author} {\bibfnamefont {R.}~\bibnamefont
  {Schinke}}\ and\ \bibinfo {author} {\bibfnamefont {P.}~\bibnamefont
  {Fleurat-Lessard}},\ }\href {\doibase http://dx.doi.org/10.1063/1.1784776}
  {\bibfield  {journal} {\bibinfo  {journal} {The Journal of Chemical Physics}\
  }\textbf {\bibinfo {volume} {121}},\ \bibinfo {pages} {5789} (\bibinfo {year}
  {2004})}\BibitemShut {NoStop}%
\bibitem [{\citenamefont {Whitfield}\ \emph {et~al.}(2011)\citenamefont
  {Whitfield}, \citenamefont {Biamonte},\ and\ \citenamefont
  {Aspuru-Guzik}}]{whitfield2011}%
  \BibitemOpen
  \bibfield  {author} {\bibinfo {author} {\bibfnamefont {J.~D.}\ \bibnamefont
  {Whitfield}}, \bibinfo {author} {\bibfnamefont {J.}~\bibnamefont {Biamonte}},
  \ and\ \bibinfo {author} {\bibfnamefont {A.}~\bibnamefont {Aspuru-Guzik}},\
  }\href {\doibase 10.1080/00268976.2011.552441} {\bibfield  {journal}
  {\bibinfo  {journal} {Molecular Physics}\ }\textbf {\bibinfo {volume}
  {109}},\ \bibinfo {pages} {735} (\bibinfo {year} {2011})}\BibitemShut
  {NoStop}%
\bibitem [{\citenamefont {Giovannetti}\ \emph {et~al.}(2008)\citenamefont
  {Giovannetti}, \citenamefont {Lloyd},\ and\ \citenamefont {Maccone}}]{GLM08}%
  \BibitemOpen
  \bibfield  {author} {\bibinfo {author} {\bibfnamefont {V.}~\bibnamefont
  {Giovannetti}}, \bibinfo {author} {\bibfnamefont {S.}~\bibnamefont {Lloyd}},
  \ and\ \bibinfo {author} {\bibfnamefont {L.}~\bibnamefont {Maccone}},\
  }\href@noop {} {\bibfield  {journal} {\bibinfo  {journal} {Physical review
  letters}\ }\textbf {\bibinfo {volume} {100}},\ \bibinfo {pages} {160501}
  (\bibinfo {year} {2008})}\BibitemShut {NoStop}%
\end{thebibliography}%

\end{document}